\documentclass[preprint,showpacs,preprintnumbers,amsmath,amssymb,superscriptaddress]{revtex4}

\usepackage{graphicx}
\usepackage{dcolumn}
\usepackage{bm}

\begin{document}

%
\title{Orientational relaxation  in a dispersive dynamic medium : Generalization of the Kubo-Ivanov-Anderson jump diffusion model to include fractional environmental dynamics
}
\author{
K. Seki}
\affiliation{
National Institute of Advanced Industrial Science and Technology (AIST)\\
AIST Tsukuba Central 5, Higashi 1-1-1, Tsukuba, Ibaraki, Japan, 305-8565
}
\author{
B. Bagchi}
\affiliation{
National Institute of Advanced Industrial Science and Technology (AIST)\\
AIST Tsukuba Central 5, Higashi 1-1-1, Tsukuba, Ibaraki, Japan, 305-8565
}
\affiliation{ Permanent address: Solid State and Structural Chemistry Unit, Indian Institute of Science, Bangalore 560012, India.
}
\author{M. Tachiya
}
%
\affiliation{
National Institute of Advanced Industrial Science and Technology (AIST)\\
AIST Tsukuba Central 5, Higashi 1-1-1, Tsukuba, Ibaraki, Japan, 305-8565
}

\begin{abstract}
Ivanov-Anderson (IA) model (and an earlier treatment by Kubo) envisages a decay of  the orientational 
correlation by random but large amplitude molecular jumps, as opposed to infinitesimal small jumps assumed 
in Brownian diffusion. Recent computer simulation studies on water and supercooled liquids have shown 
that large amplitude motions may indeed be more of a rule than exception. Existing theoretical studies on 
jump diffusion mostly assume an exponential (Poissonian)  waiting time distribution for jumps, thereby again leading to 
an exponential decay. Here we extend the existing formalism of Ivanov and Anderson to include an algebraic 
waiting time distribution between two jumps. 
As a result, 
the first ($\ell=1$) and second ($\ell=2$) rank orientational time correlation functions 
show the same long time power law, 
but their short time decay behavior is quite different. 
The predicted Cole-Cole plot of dielectric 
relaxation reproduces various features of non-Debye behaviour observed experimentally. We also developed a
theory where both unrestricted small jumps and large angular jumps coexist simultaneously. The small jumps are
shown to have a large effect on the long time decay, particularly in mitigating the effects of algebraic waiting 
time distribution, and in giving rise to an exponential-like decay, with a time constant, surprisingly, less than
the time constant that arises from small amplitude decay alone.

\end{abstract}
\pacs{ 05.40+jc, 64.70.Pf, 66.20.+d,  66.10.-x}
\maketitle

\newpage
\setcounter{equation}{0}
\section{Introduction}
\vspace{0.5cm}

Because orientational relaxation of  molecules is relatively easily accessed by a variety of experimental techniques (NMR, IR, fluorescence depolarization, optical Kerr effect, to name a few) \cite{Berne,Fleming,Bagchi,Debye,Cross,Fayer}, many theoretical models and microscopic studies have addressed various aspects of molecular rotation in dense liquids and glasses \cite{Fayer,Ivanov,Anderson,Madden,Chandra,Schiler}. The most celebrated of these studies is the one carried out by Debye many years ago, in terms of a simple rotational diffusion equation.\cite{Bagchi,Debye} The theory assumes that rotational correlation in liquids decay by small amplitude rotational Brownian motion, with a rotational diffusion coefficient, $D_{\rm R}$. The theory makes simple prediction that the decay of the correlation functions of all ranks of Legendre function is exponential and is given by
\begin{equation}
C_\ell (t) \equiv \langle P_\ell \left[ \cos \theta (t)  \right] \rangle = \exp(- \ell(\ell+1)D_{\rm R} t) .
\label{Legendre}
\end{equation}
 The Debye model of rotational diffusion has played a pivotal role in most of the discussions on rotational diffusion in the condensed phases. Many theoretical studies have attempted to improve upon Debye model,  using  for example,  a generalized Langevin equation approach which ultimately leads to a time or frequency dependent diffusion coefficient. But all these approaches use the general assumption of infinitesimal rotation of Brownian motion.

Recent experimental and theoretical studies on supercooled liquids and surprisingly, liquid water, have shown a marked departure from the classical Debye behaviour. \cite{Leporini,Hynes} In these cases, rotational diffusion is found to occur by large amplitude jumps. In supercooled liquids and glasses, the orientational relaxation is often markedly non-exponential. \cite{Leporini} 
In liquid water, the non-exponentiality, at room temperature, is weak but relaxation becomes progressively non-exponential at low temperatures. \cite{Hynes} 
The rotational relaxation (and also translational diffusion) seems to occur primarily through rare but large amplitude jumps.  A quantitative understanding of the origin of such jumps has remained a subject of great interest, though largely unsolved.

A model of rotational jump diffusion was actually proposed by several workers in the past, 
most notably by Kubo \cite{Kubo,Kubobook} and Ivanov \cite{Ivanov}.  
In Kubo's model of jump diffusion, the rotator was restricted to jump in a circle, that is restricted to two dimension. 
(See Figure 1(a)) Ivanov's model was more general where jumps were isotropically distributed in three dimension.
(See Figure 1(b))
 Ivanov's model has found increasing use in describing experimental results. In this model, the waiting time between jumps obey an exponential distribution and as a result, the decay is single exponential, with the first and second rank correlation functions which are given by
\begin{eqnarray}
C_1 (t) &=& \exp \left[-(1-\cos \Delta)t/\tau \right], 
\label{Ivanov1} \\
C_2 (t) &=& \exp \left[-\frac{3}{2}(1-\cos^2 \Delta)t/\tau \right] ,
\label{Ivanov2} 
\end{eqnarray}
where $\Delta$ is the constant amplitude of jump and $\tau$ is the average time interval between any two jumps. The probability that there are n jumps in a time interval t is assumed to be given by Poisson distribution
\begin{equation}
P(n,t) = \frac{\left(t/\tau \right)^n}{n!} \exp \left( - t/\tau \right) .
\end{equation}
For small jumps, equations 2 and 3 go over to the Debye behaviour, with $\tau_1/\tau_2=3$, 
where $\tau_1$ and $\tau_2$ are the decay time constants of $C_1(t)$ and $C_2 (t)$, respectively. 
However, the relaxation pattern is different for long jumps. The difference is most acute when the jump angle $\Delta$ is close to $\pi$. 
A jump by $\Delta \sim \pi$ relaxes $C_1(t)$ but not $C_2(t)$. therefore, $\tau_1/\tau_2$ becomes much smaller than 3 and approaches zero.  Interestingly, for intermediate values of jump length parameter ($\Delta \approx \pi/2$), the ratio of the two time constants approach unity. Simulations have often found that the values of the ratio $\tau_1/\tau_2$ lies close to unity in supercooled liquids which has been taken as an indication of the emergence of jump diffusion as a contributing mechanism of rotational relaxation.
On the other hand, in a notable development, recently D\'ejardin and Jadzyn have extended the Debye rotational diffusion model to  fractional rotational diffusion case to treat events that are non-local in time due to the existence of extensive memory effects. \cite{Dejardin} 
Under fractional rotational Brownian motion, the decay of the relevant orientational correlation functions is no longer exponential (or, sum of a few exponential terms). It was shown that the decay can be described as a combination of Mittag-Leffler temporal pattern, behaving like a stretched exponential at short times and an inverse power law in the long time limit.  Fractional equations provide an anomalously slow decay at long times, often referred to as subdiffusive regime. As pointed out by many, fractional Fokker-Planck or Smoluchowski equation is a natural generalization of normal diffusion to disordered systems with scale free memory effects. \cite{BarkaiPRE,BarkaiJPC,Sokolov02,Metzler04}
D\'ejardin and Jadzyn obtained analytic expressions of the frequency ($\omega$) dependent electric birefringence spectrum, $\chi(\omega)$. They plotted Cole-Cole diagram for various cases to demonstrate the effects of fractional diffusion. Fractional diffusion gives rise to markedly non-Debye Cole-Cole plot which now varies from 
  Cole-Davidson skewed arc behaviour  to Cole-Cole depressed circle.
The treatment of D\'ejardin and Jadzyn is still based on rotational diffusion equation, and does not contain the effects of finite jumps.  However, as already discussed, in many complex systems, rotational relaxation occurs by large amplitude jumps. For example, in liquid water, it has recently been discussed how much of the relaxation of orientational correlation occurs by jumps which are of the order of $60^\circ$. \cite{Hynes} 
Such motions cannot be treated as a Brownian diffusion.  In addition, in  many complex systems, the waiting time distribution between jumps may not be  approximated by Poissonian distribution. \cite{Leporini,BarkaiPRE,BarkaiJPC,Sokolov02,Metzler04,Klafter,SBT}
In the present work we have extended the theory of Kubo, Ivanov and Anderson to treat jump diffusion with an algebraic waiting time distribution. Since the treatments of Kubo are different from that of Ivanov and Anderson, separate solutions have been obtained. As is usually the case for relaxation with fractional diffusion,  an analytical  solution of the orientational time correlation function has been  obtained only in the frequency domain. However, we have been able to obtain asymptotic solution in all the cases. 
The decay of the correlation of the first and the second rank harmonic follow power law.  This signifies a breakdown of Debye behaviour in dispersive medium.
We have also developed a theory where the large amplitude jumps simultaneously coexist with 
small amplitude jumps. 
Interestingly, the exponential kinetics is recovered even under the small fraction of small 
amplitude jumps but 
the decay is accelerated by large amplitude jumps.

\section{\label{sec: model} Generalization of jump model}
\vspace{0.5cm}

\subsection{Kubo model}
We consider a 2-dimensional rotator which makes the series of jumps.  
The model was originally proposed by Kubo for rotational relaxation of spin 
under the presence of pulsed magnetic fields. \cite{Kubo,Kubobook}
The angle changes by the jump with the amount, $\Delta_n$. 
According to Kubo we define, 
\begin{eqnarray}
f (t) &=& \left\langle  \sum_{n=0}^\infty \exp \left( i  \sum_{m=0}^n \Delta_m  \right) P(n,t) \right\rangle , 
\end{eqnarray}
where 
$P(n,t)$ represents the probability of having $n$ jumps up to time $t$ and 
$\langle \cdots \rangle$ denotes the ensemble average of scattering angles. 
The correlation functions are obtained by, 
\begin{eqnarray}
C_\ell (t) &=& \mbox{Re} \left[ f^\ell (t) \right] ,
\label{CKubo}
\end{eqnarray}
where $\ell=1,2$.  
It should be noticed here that the $\ell-$th rank correlation function is defined in terms of $\cos \ell \theta(t)$ 
for two dimensional rotator. 
Even when the second rank correlation function is defined in terms of Legendre function of  
three dimensional isotropic rotator, 
the right hand side of eq. (\ref{CKubo}) is equal to 
$\left(C_\ell (t) - C_\ell (\infty)\right)/\left(C_\ell (0) - C_\ell (\infty)\right)$, 
which still expresses the decay of correlation function. 
We calculate $\mbox{Re} \left[ f^\ell (t) \right] $ and represents it by $C_\ell (t)$ 
of 2 dimensional rotator for Kubo model. 
The waiting time distribution of each time interval $\tau_n = t_n - t_{n-1}$, is assumed to be 
statistically independent and it is represented by, 
$\psi (\tau)$.  
We also introduce 
$\varphi (t) = \int_t^\infty d t_1 \psi (t_1)$ 
as a probability 
that a molecular rotor will not make a jump 
for the time interval between $0$ and $t$. 
The probability of having $n$ jumps up to time $t$ is given 
after the Laplace transformation, 
$\hat{P}(n,s) = \int_0^\infty \exp (-st) P (n,t)$, as \cite{Montroll65}  
\begin{eqnarray}
\hat{P}(n,s) = \hat{\varphi} (s) \hat{\psi}^n (s), 
\label{eq2}
\end{eqnarray} 
where $\hat{\varphi} (s)= (1- \hat{\psi} (s))/s$. 
The amount of each jump is also assumed to be statistically independent, 
\begin{eqnarray}
\langle \exp ( i \sum_n \Delta_n ) \rangle = \exp (i n \Delta)  .
\end{eqnarray}
Then, we have, 
\begin{eqnarray}
C_\ell (t)  = \mbox{Re} \left[ \sum_{n=0}^\infty \exp (i \ell n \Delta) P (n,t) \right]. 
\label{eq1}
\end{eqnarray}
By substituting eq. (\ref{eq2}) into eq. (\ref{eq1}), 
the Laplace transform of it becomes, 
\begin{eqnarray}
\hat{C}^\ell (s)  = \frac{1- \hat{\psi} (s)}{s} \mbox{Re} \left[
\frac{1}{1- \exp (i \ell \Delta)  \hat{\psi} (s)} 
\right].
\label{beautyKubo}
\end{eqnarray}
The above equation holds for an arbitrary waiting time distribution function and an arbitrary jump amplitude. 
When the waiting time distribution decays exponentially with the characteristic time of the jump interval, $\tau$, 
the well known result for the Poisson noise is recovered, {\it i.e.}, 
\begin{eqnarray} 
C_\ell (t)  = \exp \left[ -(t/\tau) \left( 1- \cos(\ell  \Delta )  \right) \right] 
\cos\left[(t/\tau) \sin (\ell  \Delta ) \right] . 
\label{Kubo}
\end{eqnarray}
Eq. (\ref{beautyKubo}) is the generalization of Kubo model 
to the case of an arbitrary waiting time distribution. 
Eq. (\ref{Kubo}) was given by Kubo for the Poissonian waiting time distribution for $\ell=1$.  \cite{Kubo,Kubobook}
In 2 dimensional rotational diffusion, 
eq. (\ref{Legendre}) is replaced by $\exp \left( -\ell^2 D_R t \right)$ \cite{Zwanzigbook}
since the correlation function is defined by $\langle \cos \left( \ell \theta (t) \right) \rangle$. 
Eq. (\ref{Kubo}) gives $\tau_1/\tau_2 =4$ in the limit of small amplitude jump, 
which is consistent with the result of conventional 2-dimensional rotator. 

\subsection{Ivanov-Anderson model}

For dielectric relaxation, however, Ivanov-Anderson model is more popular to describe 
the random scattering of angle. \cite{Ivanov,Anderson}
Ivanov-Anderson model is, however, similar in spirit to Kubo model. 
Ivanov-Anderson model assumes for simplicity an isotropic reorientation by random angular jumps. 
If we denote the angular change by $n$-th jump by   
$\Delta_{n}$ and assume that 
it is statistically independent as in Kubo model,  we have, 
\begin{eqnarray}
L_1&=& \cos \Delta= \langle \cos \Delta_{n} \rangle, \\
L_2 &=& (3\cos^2 \Delta-1)/2= 
\langle \left[ \left(3 \cos^2 \Delta_{n} -1 \right) \right]/2 \rangle .
\end{eqnarray}
It is known that the correlation function is expressed in the form similar to eq. (\ref{eq1}) as, 
\begin{eqnarray}
C_\ell (t) = \sum_{n=0}^\infty L_\ell^n P (n,t).  
\label{eqc1}
\end{eqnarray}
By substituting the Laplace transform of eq. (\ref{eq2}), 
the Laplace transform of the correlation function,  eq. (\ref{eqc1}), becomes, 
\begin{eqnarray}
\hat{C}_\ell (s) = \frac{1- \hat{\psi} (s)}{s} 
\frac{1}{1- L_\ell \hat{\psi} (s)} .
\label{beautyIvanov}
\end{eqnarray}
Eq. (\ref{beautyIvanov}) is valid for an arbitrary waiting time distribution. 
When the waiting time distribution between jumps decays exponentially with the characteristic time $\tau$, 
eq. (\ref{beautyIvanov}) recovers the well known results of eqs. (\ref{Ivanov1})-(\ref{Ivanov2}).
Here, the Ivanov-Anderson model is generalized to the case of an arbitrary waiting time distribution. 

\section{\label{sec:AWT} Algebraic waiting time distribution of jump}

Now, we study the influence of the power law waiting time distribution $\psi (t) \sim 1/ t^{\alpha+1}$ 
on the correlation functions. 
A well-known and popular waiting time distribution function which is normalizable is given by, 
\begin{eqnarray}
\psi (t) &=& \frac{\alpha \gamma \left( \alpha + 1, \gamma_{\rm r} t\right)}{\gamma_{\rm r}^\alpha t^{\alpha+1}} , 
\label{eq3}
\end{eqnarray}
where  
$
\gamma (z, p) \equiv \int_0^p e^{-t} t^{z-1} d\,t \mbox{  for } (\mbox{Re} z > 0)
$
is the incomplete Gamma function.  \cite{Abramowitz}
Eq. (\ref{eq3}) can be derived from 
a model in which the hopping rate depends on a parameter exponentially 
(for example activation energy) and 
the value of this parameter 
has an exponential distribution. \cite{Tachiya75}
$\gamma_{\rm r}$ is a parameter characterizing the hopping frequency. 
The waiting time distribution of eq. (\ref{eq3}) can be written approximately in a more transparent form, 
\begin{eqnarray}
\psi (t)  
\approx \frac{\alpha \Gamma \left( \alpha + 1\right)}{\gamma_{\rm r}^\alpha t^{\alpha+1}}, 
\label{eq4}
\end{eqnarray}
where $\Gamma (z)$ is the Gamma function. \cite{Abramowitz} 
The Laplace transform of the waiting time distribution is obtained as,
\begin{eqnarray}
\hat{\psi} (s) &=& 1 - \,_2F_1 \left[1, \alpha, \alpha+1, - \gamma_{\rm r} /s \right] . 
\label{WTDhyper}
\end{eqnarray}
Eq.  (\ref{WTDhyper}) is useful because 
the following property of the hypergeometric function
is known \cite{Abramowitz},
\begin{eqnarray}
\,_2F_1 \left[1, \alpha, \alpha+1, - \gamma_{\rm r} /s \right]
\approx  
\displaystyle \frac{\pi \alpha}{\sin \pi \alpha} \left( \frac{s}{\gamma_{\rm r}} \right)^{\alpha} \hspace{1cm} \displaystyle \frac{s}{\gamma_{\rm r}} < 1 .
\label{wtdLa}
\end{eqnarray}
By substituting eq. (\ref{WTDhyper}) into eq. (\ref{beautyKubo}), extended Kubo model leads to, 
\begin{eqnarray}
\hat{C}_\ell (s) &\approx& \frac{\,_2F_1 \left[1, \alpha, \alpha+1, - \gamma_{\rm r} /s \right]}{s} 
\mbox{Re} \left( \frac{1}{1- \exp (i \ell \Delta) } \right)\\
&\approx& \frac{\pi \alpha}{\sin \pi \alpha} \left[
\frac{1-\cos(\ell \Delta)}{\left(1-\cos(\ell \Delta)\right)^2 + \sin^2 (\ell \Delta)} 
\right]
\frac{1}{s^{1-\alpha} \gamma_{\rm r}^\alpha}.  
\end{eqnarray}
By performing the inverse Laplace transformation, we find the asymptotic time dependence for 
extended Kubo model, 
\begin{eqnarray}
C_\ell (t) \approx \frac{\pi \alpha}{\sin \pi \alpha} \left[
\frac{1-\cos(\ell \Delta)}{\left(1-\cos(\ell \Delta)\right)^2 + \sin^2 (\ell \Delta)} 
\right]
 \frac{1}{\Gamma(1-\alpha) 
\left( \gamma_{\rm r} t \right)^\alpha} . 
\end{eqnarray}

We can obtain the solution of the extended Ivanov-Anderson model 
in a similar fashion. 
By substituting eq. (\ref{wtdLa}) into eq. (\ref{WTDhyper}), 
eq. (\ref{beautyIvanov}) becomes, 
\begin{eqnarray}
\hat{C}_\ell (s) = \frac{1}{L_\ell} 
\frac{s^{\alpha-1}}{
\displaystyle s^\alpha + 
\frac{1-L_\ell}{L_\ell} \frac{\sin \pi \alpha}{\pi \alpha} \gamma_{\rm r}^\alpha
}. 
\label{Cls_Ivanov}
\end{eqnarray}
In the long time limit, 
\begin{eqnarray}
\left(\gamma_{\rm r} t \right)^\alpha
> \frac{L_\ell}{1-L_\ell}
\frac{\pi \alpha} {\sin (\pi \alpha)}, 
\label{longtime}
\end{eqnarray}
we find, in the small $s$ limit, the following asymptotic result, 
\begin{eqnarray}
\hat{C}_\ell (s) &\approx& \frac{\pi \alpha}{\sin \pi \alpha} \frac{1}{s^{1-\alpha} \gamma_{\rm r}^\alpha}
\frac{1}{1- L_\ell} . 
\end{eqnarray}
By performing the inverse Laplace transformation, we find the asymptotic time dependence of 
extended Ivanov-Anderson model as, 
\begin{eqnarray}
C_\ell (t) \approx \frac{\pi \alpha}{\sin \pi \alpha} \frac{1}{(1- L_\ell)}  \frac{1}{\Gamma(1-\alpha) 
\left( \gamma_{\rm r} t \right)^\alpha} . 
\label{algt}
\end{eqnarray}
Both Kubo and Ivanov-Anderson models predict exactly the same time dependence of the correlation functions, 
which is described by the algebraic time dependence with the exponent $\alpha$. 
As we can see from eq. (\ref{longtime}), 
the above asymptotic dependence is hardly attained 
when $L_\ell \approx 1$.  
This case can be examined more rigorously. 
By the inverse Laplace transformation of eq. (\ref{Cls_Ivanov}), 
we find the correlation functions expressed in terms of Mittag-Leffler function, 
\begin{eqnarray}
C_\ell (t) = 
E_\alpha 
\left[ - \frac{1-L_\ell}{L_\ell} 
\frac{\sin \pi \alpha}{\pi \alpha} 
\left(\gamma_{\rm r} t \right)^\alpha \right], 
\label{sML}
\end{eqnarray}
where the Mittag-Leffler function is defined by, 
$E_\alpha (z) = \sum_0^\infty z^k/\Gamma(\alpha k + 1)$ 
and $L_\ell \approx 1$ is introduced. 
Mittag-Leffler function is approximated by the stretched exponential function, 
\begin{eqnarray}
C_\ell (t) \approx \exp 
\left[ - 
\frac{1-L_\ell}{\Gamma(1+ \alpha) L_\ell}
\frac{\sin (\pi \alpha)}{\pi \alpha} 
\left(\gamma_{\rm r} t \right)^\alpha \right], 
\label{strexp}
\end{eqnarray}
except the final component showing algebraic decay of 
eq. (\ref{algt}) which appears at very long times when $L_\ell \approx 1$. 
$L_\ell \approx 1$ occurs 
when $\Delta \simeq 0$ for both $\ell=1$ and 
$\ell=2$. 
When $\Delta \simeq \pi$, 
$L_2 \approx 1$ but $L_1 < 1$. 
Therefore, when $\Delta$ is small, 
both $C_1(t)$ and $C_2 (t)$ mainly decay according to 
stretched exponential law. 
As $\Delta$ is increased into the range, 
$0 \ll \Delta \ll \pi$, 
both $C_1 (t)$ and $C_2 (t)$ exhibit algebraic decay of eq. (\ref{algt}). 
When $\Delta$ is closed to $\pi$, 
$C_2(t$) mainly decays by the stretched exponential law and 
the final small components of  
both $C_1 (t)$ and $C_2 (t)$ are described by algebraic decay.

By the known identity of the Mittag-Leffler function for $\alpha=0.5$, 
eq. (\ref{sML}) can be expressed as, 
\begin{eqnarray}
C_\ell (t) = 
\exp \left[ 
\left( \frac{2 \left( 1-L_\ell \right)}{\pi L_\ell}  \right)^2
\gamma_{\rm r} t  \right] 
\mbox{erfc} \left[ \frac{2\left( 1-L_\ell \right)}{\pi L_\ell}  
\sqrt{\gamma_{\rm r} t} \right] , 
\end{eqnarray}
for $\alpha=1/2$, 
where the complementary error function is defined by $\mbox{erfc} (z) = (2/\sqrt{\pi}) \int_z^\infty \exp (- y^2 ) dy$. \cite{Abramowitz} 
We compare these analytical results with the numerical Laplace inversion of the exact results 
(eqs. (\ref{beautyKubo}) and (\ref{beautyIvanov}) with eq. (\ref{WTDhyper})) in section \ref{sec:results}. 

\newpage
\section{\label{sec:jumps} Simultaneous Coexistence of Large and small amplitude jumps}

In many situations, jumps with large amplitudes are rare and they are superimposed on 
the small amplitude jumps which occur frequently. 
We denote the waiting time distribution of large amplitude jump by, $\psi_a (t)$ and that of 
small amplitude jump by $\psi_b (t)$. 
Theoretically, they are related with each other. 
By denoting the waiting time distribution of large amplitude jump in the absence of small amplitude jump by, $\psi^{(0)}_a (t)$, 
and that of small amplitude jump in the absence of large amplitude jump by, $\psi^{(0)}_b (t)$, 
the waiting time distribution under the presence of both types of jumps is expressed as, \cite{Barzykin}
\begin{eqnarray}
\psi_a(t) &=& \psi^{(0)}_a (t) \left(1- \int_0^t \, dt \psi^{(0)}_b (t) \right),\\
\psi_b(t) &=& \psi^{(0)}_b (t) \left(1- \int_0^t  \, dt \psi^{(0)}_a (t) \right).  
\end{eqnarray}
since the large amplitude jump occurs before the occurrence of small amplitude jump and vice versa. 

First we consider the case of Kubo model. 
The Laplace transform of the correlation function is generalized from eq. (\ref{beautyKubo}) as, 
\begin{eqnarray}
\hat{C}_\ell (s) &=& 
\frac{1- \hat{\psi}_a(s)- \hat{\psi}_b(s) }{s}
\mbox{Re} \left[ 
\sum_{n=0}^\infty \left( 
\exp\left( i \ell \Delta_a \right)  \hat{\psi}_a (s) +
\exp\left( i \ell \Delta_b \right) \hat{\psi}_b (s) 
\right)^n
\right] \\
&=& \frac{1- \hat{\psi}_a(s)- \hat{\psi}_b(s) }{s}
\mbox{Re} \left[ 
\frac{1}
{1-  \exp\left( i \ell \Delta_a \right)  \hat{\psi}_a (s) -
\exp\left( i \ell \Delta_b \right) \hat{\psi}_b (s) }
\right] , 
\label{beautyKubo2}
\end{eqnarray}
where we assume that each jump belonging to either large or small amplitude is statistically independent and 
the average is denoted by, 
$ \exp\left( i \ell \Delta_j \right)$, where $j=a,b$.

For Ivanov-Anderson  model, 
the Laplace transform of the correlation function is generalized from eq. (\ref{beautyIvanov}) as, 
\begin{eqnarray}
\hat{C}_\ell (s) &=& 
\frac{1- \hat{\psi}_a(s)- \hat{\psi}_b(s) }{s}
\left[ 
\sum_{n=0}^\infty \left( 
L_\ell^{(a)}  \hat{\psi}_a (s) +
L_\ell^{(b)}  \hat{\psi}_b (s) 
\right)^n
\right] \\
&=& \frac{1- \hat{\psi}_a(s)- \hat{\psi}_b(s) }{s}
\left[ 
\frac{1}
{1-L_\ell^{(a)}   \hat{\psi}_a (s) -
L_\ell^{(b)}  \hat{\psi}_b (s) }
\right] , 
\label{beautyIvanov2}
\end{eqnarray}
where we again assume that each jump belonging to either large or small amplitude is statistically independent and 
the average is denoted by, 
\begin{eqnarray}
L_1^{(j)} &=& \cos \Delta_j , \\
L_2^{(j)} &=& \left(3 \cos^2 \Delta_j -1 \right) /2 ,
\end{eqnarray}
 where $j=a,b$.

When the both amplitude jumps occur according to the exponential waiting time distribution, 
\begin{eqnarray}
\psi_a (t) &=& \frac{1}{\tau_a} \exp \left( - t/\tau_a -t/\tau_b \right) \\
\psi_b (t) &=& \frac{1}{\tau_b} \exp \left( - t/\tau_a -t/\tau_b \right) , 
\end{eqnarray}
Kubo's result of eq. (\ref{Kubo}) is generalized to, 
\begin{eqnarray}
C_\ell (t)   = \exp \left[ - \sum_{j=a,b} (t/\tau_j) \left( 1 - \cos (\ell \Delta_j ) \right) \right] 
\cos \left[ \sum_{j=a,b} (t/\tau_j) \sin (\ell \Delta_j) \right] .
\end{eqnarray}

While, Ivanov-Anderson's results, eqs. (\ref{Ivanov1})-(\ref{Ivanov2}) are generalized to, 
\begin{eqnarray}
C_1 (t) &=& \exp \left[ - \sum_{j=a,b} \left( 1 -\cos \Delta_j \right) (t/\tau_j) \right] ,
\label{exactIAm1}\\
C_2 (t)  &=& \exp \left[ - \sum_{j=a,b} \frac{3}{2}\left( 1 -\cos^2 \Delta_j \right) (t/\tau_j) \right]  , 
\label{exactIAm2}
\end{eqnarray}
as expected from the Markovian kinetics.

When, the large amplitude jump obeys the algebraic waiting time distribution, 
\begin{eqnarray}
\psi^{(0)}_a (t) &=& \frac{\alpha \gamma \left( \alpha+1, t/\tau_a \right)}
{ t(t/\tau_a)^{\alpha}} , 
\end{eqnarray}
and the small amplitude jump obeys the
exponential kinetics,
\begin{eqnarray}
\psi^{(0)}_b (t) &=&\frac{1}{\tau_b} \exp \left(- t/\tau_b \right)  , 
\end{eqnarray} 
the waiting time distributions under the presence of both types of jumps are given by, 
\begin{eqnarray}
\psi_a (t) &=& \psi_a^{(0)} (t ) \exp \left( - t/\tau_b \right) ,\\
\psi_b (t) &=& \frac{1}{\tau_b} \exp \left( - t/\tau_b \right) \int_t^\infty dt_1 \psi_a^{(0)} (t_1). 
\end{eqnarray}
The Laplace transform is obtained as, 
\begin{eqnarray}
\hat{\psi}_a (s) &=& \hat{\psi}_a^{(0)} (s+1/\tau_b) 
\label{psia}\\
\hat{\psi}_b (s) &=& \frac{1/\tau_b}{s+1/\tau_b} 
\left[1- \hat{\psi}_a^{(0)} (s+1/\tau_b) \right], 
\label{psib}
\end{eqnarray}
where $\hat{\psi}_a^{(0)} (z)$ is expressed as, 
\begin{eqnarray}
\hat{\psi}_a^{(0)} (z)  &=& 1- \,_2F_1 
\left[1, \alpha, \alpha+1, - 1/\left(\tau_a z \right) \right] .
\end{eqnarray}

By substituting eqs. (\ref{psia}) and (\ref{psib}) into eqs. (\ref{beautyKubo2}) and (\ref{beautyIvanov2}), 
the exact solutions are obtained in the Laplace domain. 
However, 
the inverse Laplace transform is too
complicated for analytical calculation and will be performed numerically.

Since we are interested in the case where the characteristic time of small amplitude jump is much 
shorter than that of large amplitude jump, 
the approximate solutions can be obtained by taking the limit, 
$\,_2F_1 \left[1, \alpha, \alpha+1, - 1/\left(\tau_a(s +1/\tau_b) \right)\right]
\approx \,_2F_1 \left[1, \alpha, \alpha+1, - \tau_b /\tau_a \right]$, 
which leads to  $\hat{\psi}_a^{(0)} (s+1/\tau_b) \approx \hat{\psi}_a^{(0)} (1/\tau_b)$. 
In the approximation, 
the $s$-dependence in the Laplace transform of waiting time distribution is dropped out. 
The $s$-dependence can be ignored when we are interested in the time scale larger than 
that of small amplitude jumps. 
The quantity is just the time integration over waiting time distribution of large amplitude jumps 
and represents the probability of occurrence of large amplitude jumps, {\it i.e.}  
the escape probability from small amplitude jumps by a large amplitude jump.  
By introducing the approximation, 
we find that 
the correlation functions decay exponentially, 
\begin{eqnarray}
C_1 (t) &=& \exp \left[-\left( t/\tau_b \right) 
\left( 1- \frac{\cos\left(\Delta_b\right)\hat{\psi}_b (0)}
{1- \cos \left(\Delta_a\right) \hat{\psi}_a (0)} \right)\right] ,
\label{aprxm1}\\
C_2 (t) &=& \exp \left[- 3 \left( t/\tau_b  \right)
\frac{
1- \cos(2 \Delta_a) \hat{\psi}_a (0)
-\cos(2 \Delta_b) \hat{\psi}_b (0)}
{3+\hat{\psi}_b (0)-3 \cos(2 \Delta_a) \hat{\psi}_a (0)}
\right] .
\label{aprxm2}
\end{eqnarray}
When $\Delta_a$ is close to $\Delta_b \simeq 0$, 
we recover  eqs. (\ref{Ivanov1})-(\ref{Ivanov2}) with $\Delta=\Delta_b$ and 
$\tau=\tau_b$  by substituting  
$\hat{\psi}_b (0) \approx 1$ and $\hat{\psi}_a (0) \approx 0$ into the above expressions. 
The exponential kinetics of 
eqs. (\ref{aprxm1})-(\ref{aprxm2}) 
even under the algebraic waiting time distribution of the rare but large amplitude jumps 
is the important result of this article. 

\section{\label{sec:results} Numerical results}

Fig. \ref{fig:Ivanov1} shows the relaxation of correlations of Ivanov-Anderson model for $\Delta=60^\circ$. 
For this value of $\Delta$, 
$C_2 (t)$ decays faster than $C_1 (t)$ for any values of $\alpha$. 
Compared to the decay by normal diffusion, 
decay of both $C_2 (t)$ and $C_1 (t)$ is slowed down when the waiting time distribution has 
an algebraic time dependence. 
At long times, all decay curves are well represented by 
algebraic decay of eq. (\ref{algt}) with the exponent $\alpha$. 
The long time tail in the decay of correlation originates from the algebraic time dependence of 
the waiting time distribution. 
The exponent is the same for both  $\ell=2$ and $\ell=1$ components. 

When $\Delta$ is increased to $\Delta=170^\circ$, 
$C_2 (t)$ decays slower than $C_1 (t)$ for any values of $\alpha$ 
as shown in Fig. \ref{fig:Ivanov2}. 
As $\alpha$ is decreased, the relaxation becomes slower. 
Long time asymptotic decay is again represented by the algebraic decay of eq. (\ref{algt}). 
$C_2 (t)$ decays according to Mittag-Leffler function at intermediate times 
and the fitting becomes better  
as $\alpha$ is lowered. 
Since the Mittag-Leffler function is approximated by 
the stretched exponential decay of eq. (\ref{strexp}) 
in this time range, 
$C_2 (t)$ is well fitted by the stretched exponential function. 
The stretched exponential decay appears when $\Delta$ is close to $180^\circ$ for  $C_2 (t)$ 
and $\Delta$ is close to $0^\circ$ for both  $C_2 (t)$ and $C_1 (t)$. 
For other values of $\Delta$ the analytical form of the decay is not obtained but 
as we can see from Fig. \ref{fig:Ivanov3} the decay curves are very different from the single exponential decay when 
$\alpha <1$. 
Only when $\alpha=1$, the correlations decay fast with exponential time dependence 
which is shown by the straight line in the log-linear plot. 

The results for different values of $\Delta$ with $\alpha$ kept constant are 
summarized in Fig. \ref{fig:Ivanov4}. 
For any value of $\Delta$ asymptotic time dependence is described by the algebraic time dependence with 
the exponent $\alpha$. 
The relaxation of $C_1(t)$ slows down monotonically as $\Delta$ is decreased. 
Since $C_2(t)$ is given by the square of $\cos \Delta$, 
the decay is the same when $\Delta$ is changed to $\pi-\Delta$. 
As $\Delta$ is increased from $0^\circ$, 
the decay of $C_2(t)$ becomes faster. 
The fastest decay is obtained for $\Delta=90^\circ$. 
When $\Delta$ is further increased from $\Delta=90^\circ$, 
the decay is slowed down.

In order to facilitate comparison with dielectric
relaxation experiments in dispersive medium, we investigated the 
behavior of the Cole-Cole diagram as a function of system parameters (like
jump angle and value of the power law exponent). 
The Cole-Cole diagram is obtained from the complex
dielectric function, $\epsilon (\omega)$, which satisfies, \cite{Kubobook,Zwanzig65}
\begin{eqnarray}
\frac{\epsilon (\omega) - \epsilon_\infty}{\epsilon_{\rm st} - \epsilon_\infty} 
&=& 1 + i \omega \hat{C}_\ell \left( -i \omega \right) , 
\end{eqnarray}
where $\epsilon_{\rm st} $ is the static dielectric constant and 
$\epsilon_\infty$ is the optical dielectric constant. 
The semi-circle of Debye relaxation is obtained for normal diffusion. 
As $\alpha$ is decreased, 
the maximum of $\epsilon''$ is decreased and the Cole-Cole diagram is depressed. 
When $\Delta$ is relatively small, 
the Cole-Cole diagram is symmetric, as shown in Fig. \ref{fig:Ivanov5} for any value of $\alpha$. 
Such change of Cole-Cole diagram by decreasing $\alpha$ has been  
already noticed, but without paying much attention to the influence of the jump amplitude. \cite{Metzler,Yonezawa}

Now, we investigate the influence of the jump amplitude on the Cole-Cole diagram.
For the same value of $\alpha$, 
the Cole-Cole diagram is skewed and becomes asymmetric as the jump amplitude $\Delta$ is increased 
as shown in Figs. \ref{fig:Ivanov6}-\ref{fig:Ivanov7}. 
The asymmetry is larger for smaller values of $\alpha$.

The results of Kubo model are quite similar to those of Ivanov-Anderson model. 
In Kubo model, however, 
an  oscillation is observed, in addition to the algebraic or stretched exponential decay, 
as shown in Figs. \ref{fig:Kubo1}-\ref{fig:Kubo2}. 
In Kubo model of oscillator \cite{Kubo,Kubobook} 
it has been pointed out that the jump contributes both to the damping and the oscillation 
of correlation functions. 
Dipoles rotate in a direction with an average amplitude $\Delta$, 
which gives rise to oscillation in correlation functions. 
In the case of magnetic resonance as originally studied by 
Kubo, 
the oscillation corresponds to the shift of resonant frequency 
found in the more elaborate theory. 
In the case of electric dipoles, oscillation is due to rather unphysical modeling of rotator, 
which has a preferable direction of rotation. 

The oscillation disappears by decreasing $\alpha$ values. 
When the jump amplitude is small, 
$C_1 (t)$ oscillates more than $C_2 (t)$. 
For $\Delta=60^\circ$, 
$C_1 (t)$ oscillates even when $\alpha$ is decreased to $\alpha=0.8$. 
In the case of $\Delta=170^\circ$, 
$C_2 (t)$ oscillates for normal diffusion while
other decay curves exhibit essentially monotonic decay.

So far, we have investigated the effect of long algebraic waiting time on 
the rotational relaxation with large jump amplitude. 
Large amplitude jumps are normally superimposed by small amplitude jumps. 
Therefore, we study the case when both large and small amplitude jumps are simultaneously present. 
As shown in Fig. \ref{fig:mixed1}, in this case
the kinetics is found to be almost exponential and the decay is well represented by 
eq. (\ref{aprxm1}) for $C_1(t)$ and eq. (\ref{aprxm2}) for $C_2(t)$. 
The exponential kinetics is also confirmed from the Cole-Cole diagram shown in 
Fig. \ref{fig:mixed2}, where semi-circle of Debye relaxation can be seen. 
The decay curves obtained by multiplying 
eqs. (\ref{Ivanov1})-(\ref{Ivanov2}) for small amplitude jump 
with the inverse Laplace transform of eq. (\ref{beautyIvanov}),  
where eq. (\ref{WTDhyper}) for large amplitude jump is introduced, 
are different from the exact numerical results. 
The exponential kinetics is of course different from the algebraic decay or 
stretched exponential decay found for the large amplitude jump alone. 
The small amplitude jumps are more frequent than the large amplitude jumps. 
The former gives rise to an exponential decay but 
in the presence of large amplitude jump 
the correlation functions decay almost exponentially with the time constant 
much smaller than that for the small amplitudes alone.
Thus, although the large amplitude jumps are rare, 
they contribute a lot to the relaxation. 
The waiting time distribution of large amplitude jumps in the absence of small amplitude jumps 
has an algebraic asymptotic time dependence which is easily interrupted by the more frequent small amplitude jumps.  
The decay is exponential as a result of cumulative small amplitude jumps, but
the decay is accelerated by the large amplitude jumps. 

As explained, the exponential kinetics of correlation functions, 
$C_1 (t) = \exp \left(-t/\tau_1\right)$ and $C_2 (t) = \exp \left(-t/\tau_2\right)$, 
results from the interplay  
between the frequent small amplitude jumps and 
the rare events of large amplitude jumps having algebraic waiting time distribution. 
Although the results are not shown, 
substantially the same results are obtained even when $\alpha$ is lowered to $1/2$.
For $\alpha<1$, $\tau_1$ and $\tau_2$ are obtained from 
eq. (\ref{aprxm1}) and eq. (\ref{aprxm2}), respectively, 
while  the results of normal diffusion are given by, eqs. (\ref{exactIAm1})-(\ref{exactIAm2}).

 We now investigate the jump amplitude dependence of the ratio of the correlation
times,  $\tau_1/\tau_2$.  As shown in Fig. \ref{fig:mixed3}, 
the result for $\alpha=0.8$ and that for normal diffusion are almost the same, 
although the apparent functional forms are seemingly very different. 
When $\Delta_a$ is small, the ratio is close to $3$ and it decreases monotonically to 
zero as $\Delta_a$ is increased close to $180^\circ$. 
The small deviation between the two lines is increased as 
$\Delta_a$ is increased. 
Since the results among different values of $\alpha$ are very close, 
it could be difficult to judge from the correlation functions whether  the long amplitude jumps have 
algebraic long time tail in the waiting time distribution. 

The exponential kinetics is robust as shown in Fig. \ref{fig:mixed4}. 
Even when the frequency of the small amplitude jumps is the same 
as that of the large amplitude jumps, 
the decay is still exponential. 
However, when the small amplitude jumps occur 10 times less frequently 
than the large amplitude jumps, then non-exponential kinetics is observed 
in the short time regime, $t \leq \tau_b$.

\section{\label{sec:Conclusions} Conclusions}

The models of Ivanov and of  Kubo are well-known models of the decay of  the orientational 
correlation by random but large amplitude molecular jumps. These models are expected to
give rise to decay dynamics quite distinct from the Brownian diffusion by infinitesimal small jumps.
When the waiting time distribution for large jumps is Poissonian, even the models of Ivanov and Kubo
lead to exponential decay, making the detection of appropriate molecular mechanism 
by experimental measurements rather difficult.

However, if the waiting time distribution is not Poissonian, then the difference between large jumps and the 
Brownian diffusion can be significant. In the present work, we have extended the models of Ivanov and Kubo to include
such cases. In particular, we have employed an algebraic waiting time distribution for large jumps, as
such distributions can be useful to describe supercooled liquids and glasses, and also in restricted geometries.
Recent computer simulation studies on water and supercooled liquids have indeed shown 
that large amplitude motions may be more of a rule than exception.

In the present study, we have  solved the theoretical models with algebraic
waiting time distribution analytically to obtain 
the first ($\ell=1$) and second ($\ell=2$) rank orientational time correlation functions.  As expected, 
the decay is non-exponential, with power law at longer times. 
The correlation functions for $\ell=1$ 
and   $\ell=2$  show the same long time power law 
exponents, but the short time decay behavior is quite different  for the two correlation functions. 

In order to facilitate comparison with experiments on dielectric relaxation, we have calculated Cole-Cole plots generated
by the orientational decay for a wide variety of parameters, such as jump amplitude and the power law
exponent. The predicted Cole-Cole plot of dielectric 
relaxation reproduces various features of non-Debye behaviour observed experimentally. 

In addition, We have developed a
theory where both unrestricted small jumps and large amplitude jumps coexist simultaneously. The small amplitude jumps are
shown to have a large effect on the long time decay, particularly in mitigating the effects of algebraic waiting 
time distribution. Thus, in the limit of appreciable number of small
jumps, we find the decay to become single exponential.  We find some what surprisingly that the decay time of this exponential decay is much smaller than that given by small jumps alone.

 In this work, we have not made any specific application of our theory but shown that
several features of Cole-Cole plot resemble the ones observed experimentally in complex systems. Since algebraic waiting time distribution of jumps naturally gives rise to Davidson-Cole kind of
frequency dependence of dielectric function, the present formalism should find use in interpretation of existing experimental results. Work in this direction is under progress.




\newpage
\vskip1cm
\noindent
{\bf Fig. 1:}~
a) Schematic illustration of Kubo model. 
Jumps are restricted on a circle. \\
b) Schematic illustration of Ivanov-Anderson model. 
Jumps are isotropically distributed on a sphere. 

\vskip1cm
\noindent
{\bf Fig. \ref{fig:Ivanov1}:}~
Correlation functions of Ivanov-Anderson model with $\Delta = 60^\circ$ against 
dimensionless time, $\gamma_{\rm r} t$. Red lines represent $C_2 (t)$ and black lines represent $C_1 (t)$. $\alpha=0.5$ and $0.8$ from top to bottom. Thin lines indicate the results of the normal diffusion. 
Dashed lines represent asymptotic algebraic time dependence of Eq. (\ref{algt}).

\vskip1cm
\noindent
{\bf Fig. \ref{fig:Ivanov2}:}~
Correlation functions of Ivanov-Anderson  model with $\Delta = 170^\circ$ against 
dimensionless time, $\gamma_{\rm r} t$. Red lines represent $C_2 (t)$ and black lines represent $C_1 (t)$. $\alpha=0.5$ and $0.8$ from top to bottom. Thin lines indicate the results of the normal diffusion. 
Dashed lines represent asymptotic algebraic time dependence of Eq. (\ref{algt}).
Dots represent the results of Mittag-Leffler function of Eq. (\ref{sML})

\vskip1cm
\noindent
{\bf Fig. \ref{fig:Ivanov3}:}~
Log linear plot of Fig. 2. 

\vskip1cm
\noindent
{\bf Fig. \ref{fig:Ivanov4}:}~
Correlation functions of Ivanov-Anderson  model for $\alpha=0.8$ with various values of jump amplitude against 
dimensionless time, $\gamma_{\rm r} t$. Red lines represent $C_2 (t)$ and black lines and symbols represent $C_1 (t)$.  

\vskip1cm
\noindent
{\bf Fig. \ref{fig:Ivanov5}:}~
Cole-Cole diagram of Ivanov-Anderson  model for $\Delta = 30^\circ$. 

\vskip1cm
\noindent
{\bf Fig. \ref{fig:Ivanov6}:}~
Cole-Cole diagram of Ivanov-Anderson  model for $\alpha=0.8$. 
Solid lines indicate $\Delta=10^\circ$, $60^\circ$, $90^\circ$, 
$150^\circ$, and $180^\circ$ (almost overlaps with the line of $\Delta=150^\circ$) 
 from bottom to top at $\epsilon'=0.2$. 
 Dashed line represents the Debye relaxation of $\alpha=1.0$.

\vskip1cm
\noindent
{\bf Fig. \ref{fig:Ivanov7}:}~
Cole-Cole diagram of Ivanov-Anderson  model for $\alpha=0.5$. 
Solid lines indicate $\Delta=10^\circ$, $60^\circ$, $90^\circ$, 
$150^\circ$, and $180^\circ$ (almost overlaps with the line of $\Delta=150^\circ$) 
 from bottom to top. 
 Dashed line represents the Debye relaxation of $\alpha=1.0$.

\vskip1cm
\noindent
{\bf Fig. \ref{fig:Kubo1}:}~
Correlation functions of Kubo model with $\Delta = 60^\circ$ against 
dimensionless time, $\gamma_{\rm r} t$. Red lines represent 
$C_2 (t)$ and black lines represent $C_1 (t)$. $\alpha=0.5, 0.8$ from top to bottom. Thin lines indicate the results of normal diffusion. 

\vskip1cm
\noindent
{\bf Fig. \ref{fig:Kubo2}:}~
Correlation functions of Kubo model with $\Delta = 170^\circ$ against 
dimensionless time, $\gamma_{\rm r} t$. Red lines represent 
$C_2 (t)$ and black lines represent $C_1 (t)$. $\alpha=0.5, 0.8$ from top to bottom. Thin lines indicate the results of normal diffusion. 

\vskip1cm
\noindent
{\bf Fig. \ref{fig:mixed1}:}~
Correlation functions of Ivanov-Anderson  model for $\alpha=0.8$ against 
dimensionless time, $t/\tau_{\rm a}$. Large amplitude jump of $\Delta_a=60^\circ$ 
and small amplitude jump of $\Delta_b=5^\circ$   
coexist. The frequency of small amplitude jump is 5 times larger than that of large amplitude jump, 
$\tau_a /\tau_b=5$. 
Red lines represent $C_2 (t)$ and black lines represent $C_1 (t)$.  
Thick lines indicate the exact numerical results. 
Thin lines indicate the approximate results of eqs. (\ref{aprxm1})-(\ref{aprxm2}). 
Short-dashed lines indicate the result of small amplitude jump alone. 
Long-dashed lines indicate the result of large amplitude jump alone. 
The dashed-dotted lines are the results assuming independent superposition of two processes 
as explained in the text. 

\vskip1cm
\noindent
{\bf Fig. \ref{fig:mixed2}:}~
Cole-Cole diagram of Ivanov-Anderson  model for $\alpha=0.8$. 
The solid line represents the result when large amplitude jump of $\Delta_a=60^\circ$ 
and small amplitude jump of $\Delta_b=5^\circ$   
coexist. The frequency of small amplitude jump is 5 times larger than that of large amplitude jump, 
$\tau_a /\tau_b=5$.
The dashed line indicates the result when only the large amplitude jump is present. 

\vskip1cm
\noindent
{\bf Fig. \ref{fig:mixed3}:}~
$\tau_1/\tau_2$ against the amplitude of the large amplitude jumps for the fixed jump amplitude $\Delta_b=5^\circ$
of small amplitude jumps. 
The frequency of small amplitude jump is 5 times larger than that of large amplitude jump, 
$\tau_a /\tau_b=5$. 
The solid line indicates the results of $\alpha=0.8$. 
Dots indicate the results of $\alpha=0.5$. 
The red dashed line indicates the results for the normal diffusion of  $\alpha=1.0$. 

\vskip1cm
\noindent
{\bf Fig. \ref{fig:mixed4}:}~
Correlation functions of Ivanov-Anderson  model for $\alpha=0.8$ against 
dimensionless time, $t/\tau_{\rm a}$. Large amplitude jump of $\Delta_a=60^\circ$ 
and small amplitude jump of $\Delta_b=5^\circ$   
coexist. The frequency of small amplitude jump is changed. 
Red lines represent $C_2 (t)$ and black lines represent $C_1 (t)$.  
$\tau_a /\tau_b=0.1$,$\tau_a /\tau_b=0.5$, and $\tau_a /\tau_b=1.0$ from top to bottom.

\newpage
\begin{flushright}
Fig. 1, K. Seki,  B. Bagchi, and M. Tachiya
\end{flushright}
\vspace{0cm}
\mbox{ }\\
\begin{figure}[h]
\includegraphics[width=10cm]{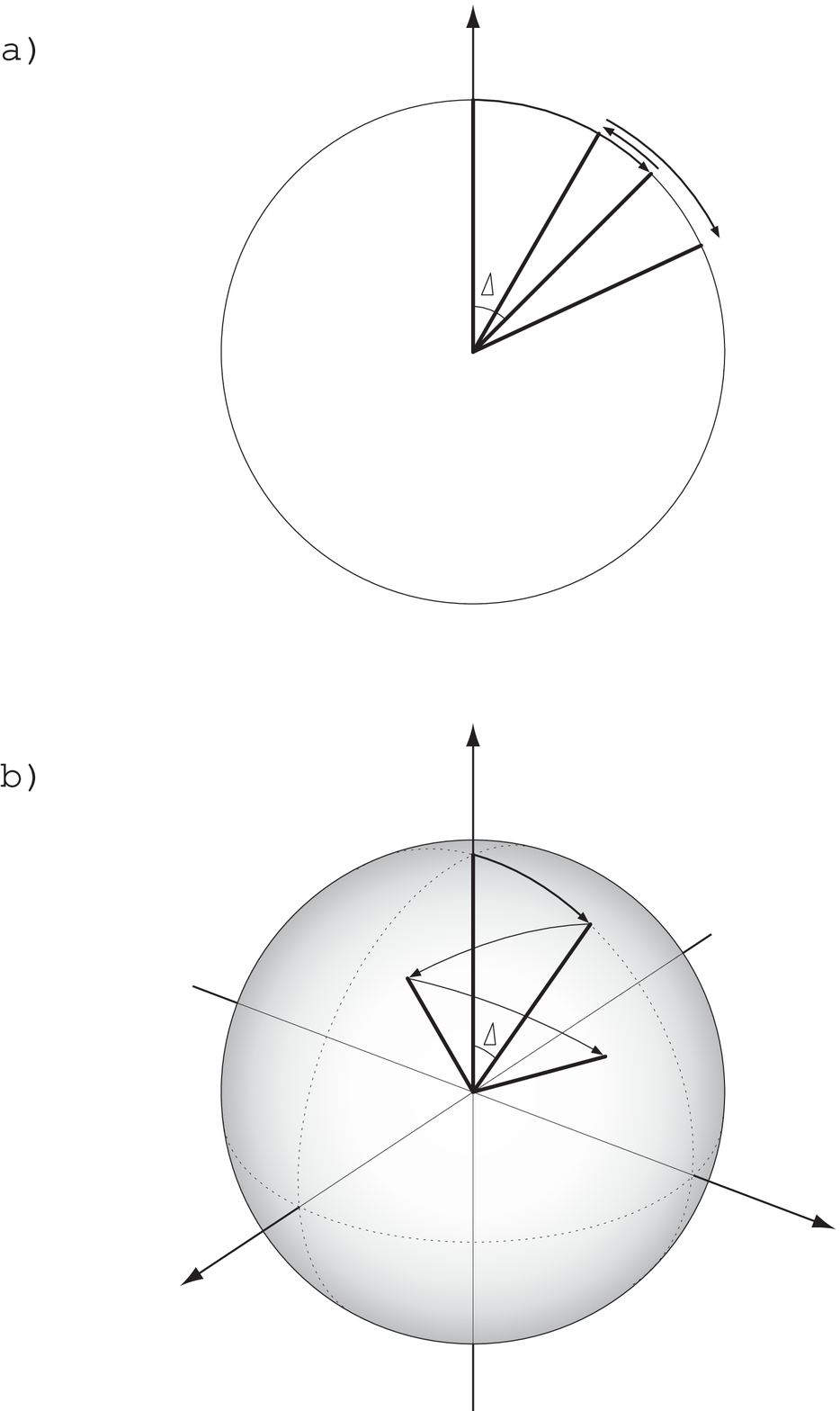}
\caption{}\label{fig:KuboIvanov} 
\end{figure}

\newpage
\begin{flushright}
Fig. 2, K. Seki,  B. Bagchi, and M. Tachiya
\end{flushright}
\vspace{5cm}
\mbox{ }\\
\begin{figure}[h]
\includegraphics[width=12cm]{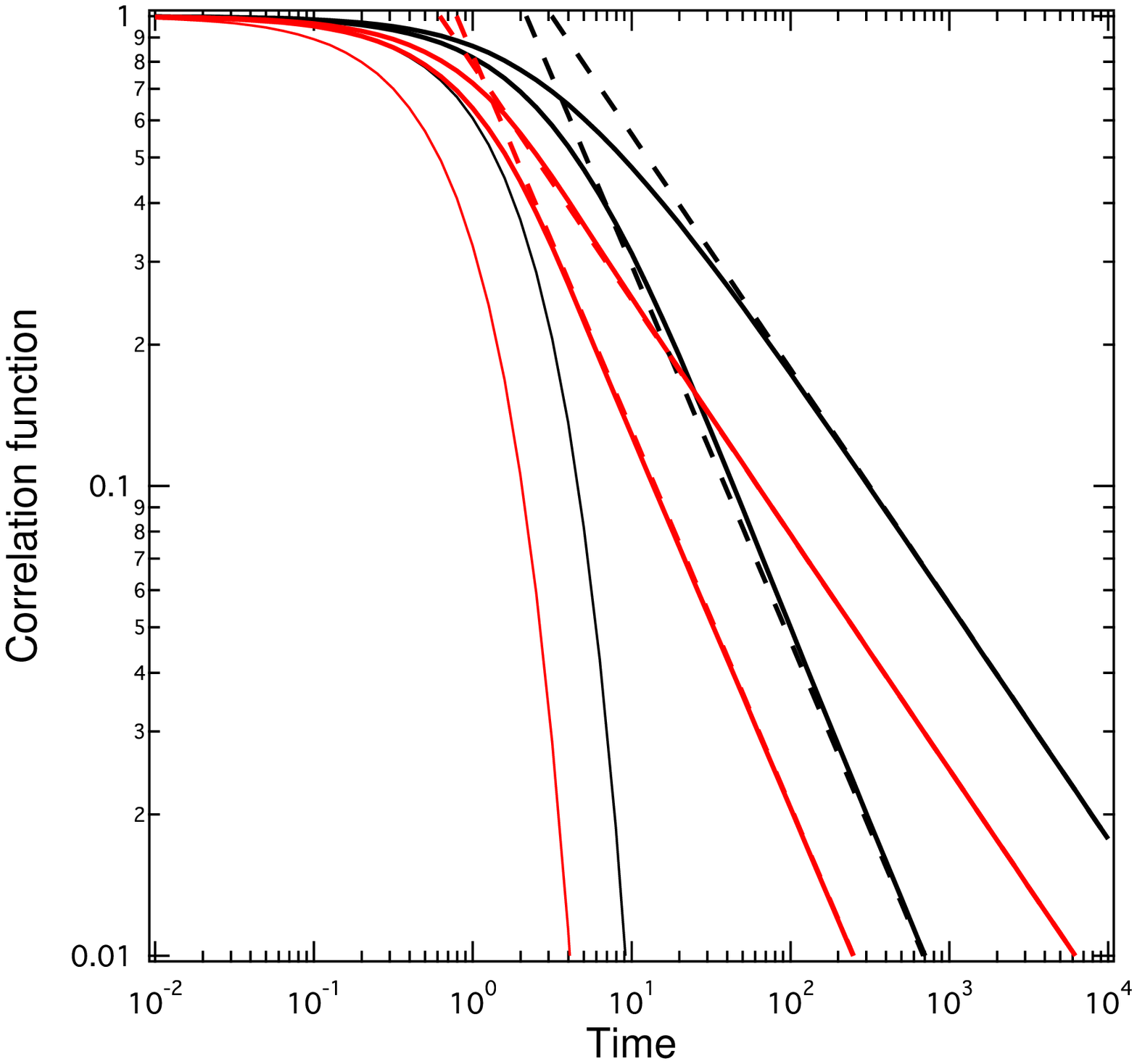}
\caption{}\label{fig:Ivanov1} 
\end{figure}

\newpage
\begin{flushright}
Fig. 3, K. Seki,  B. Bagchi, and M. Tachiya
\end{flushright}
\vspace{5cm}
\mbox{ }\\
\begin{figure}[h]
\includegraphics[width=12cm]{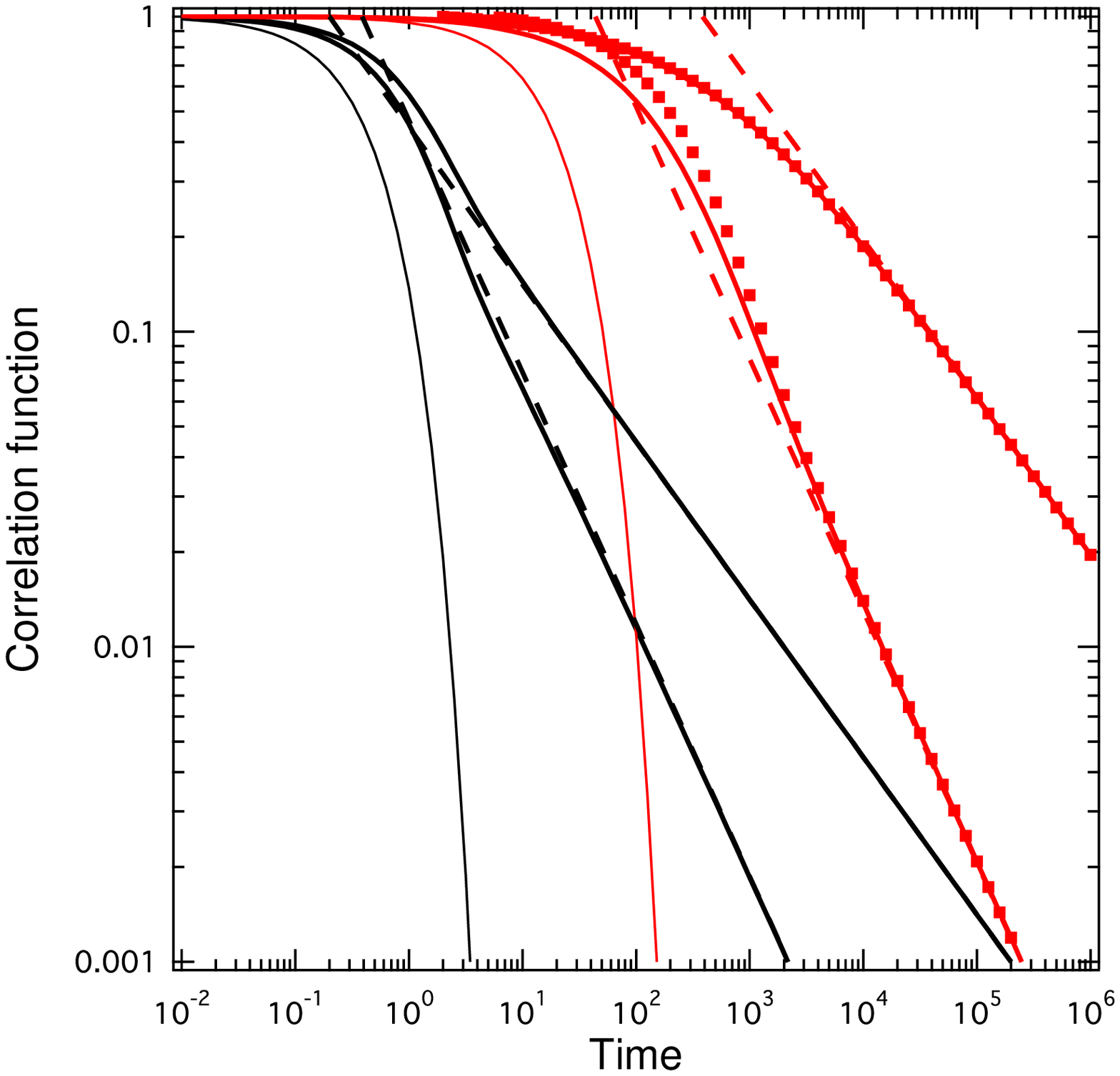}
\caption{}\label{fig:Ivanov2} 
\end{figure}

\newpage
\begin{flushright}
Fig. 4, K. Seki,  B. Bagchi, and M. Tachiya
\end{flushright}
\vspace{5cm}
\mbox{ }\\
\begin{figure}[h]
\includegraphics[width=12cm]{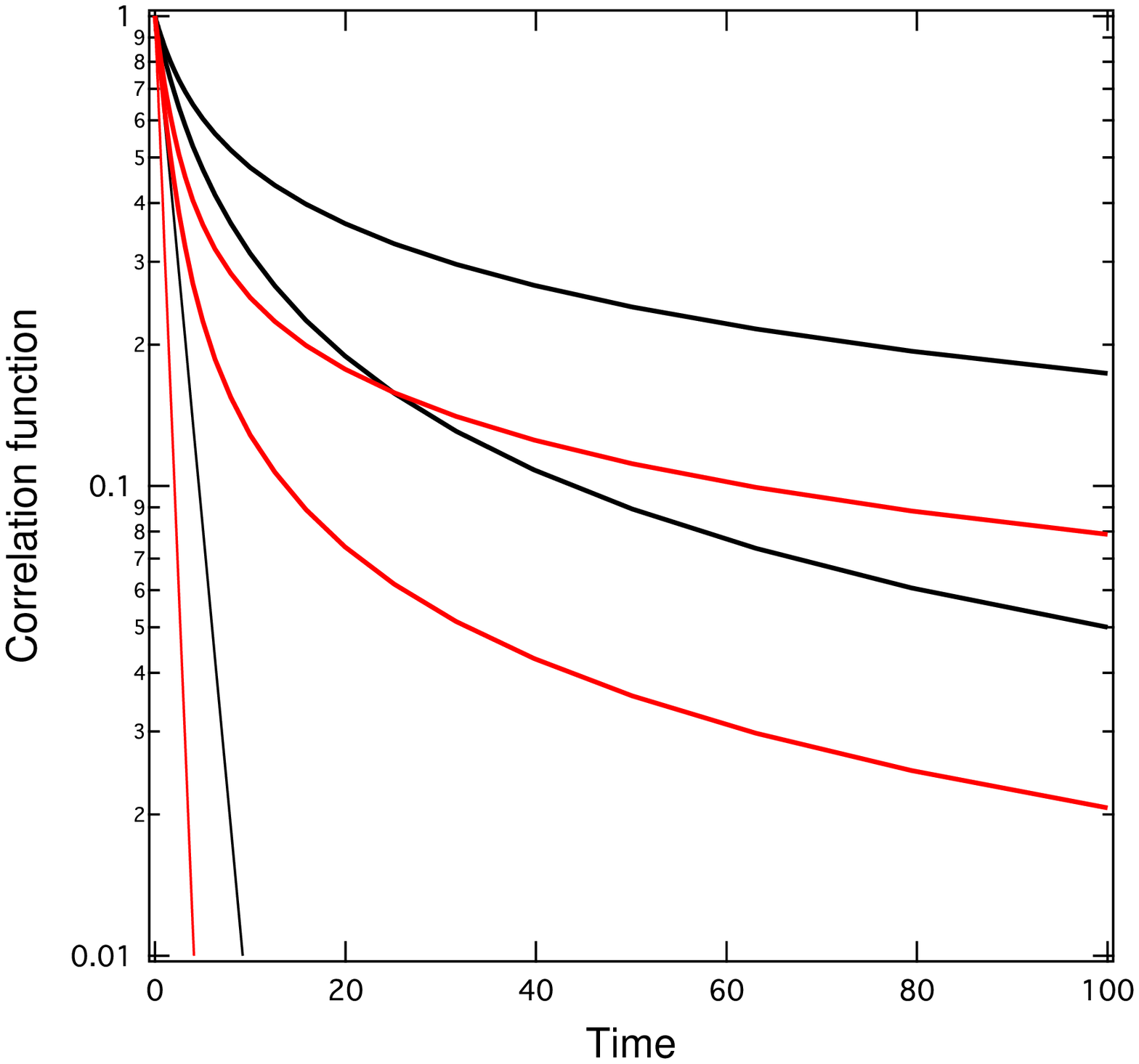}
\caption{}\label{fig:Ivanov3} 
\end{figure}

\newpage
\begin{flushright}
Fig. 5, K. Seki,  B. Bagchi, and M. Tachiya
\end{flushright}
\vspace{5cm}
\mbox{ }\\
\begin{figure}[h]
\includegraphics[width=12cm]{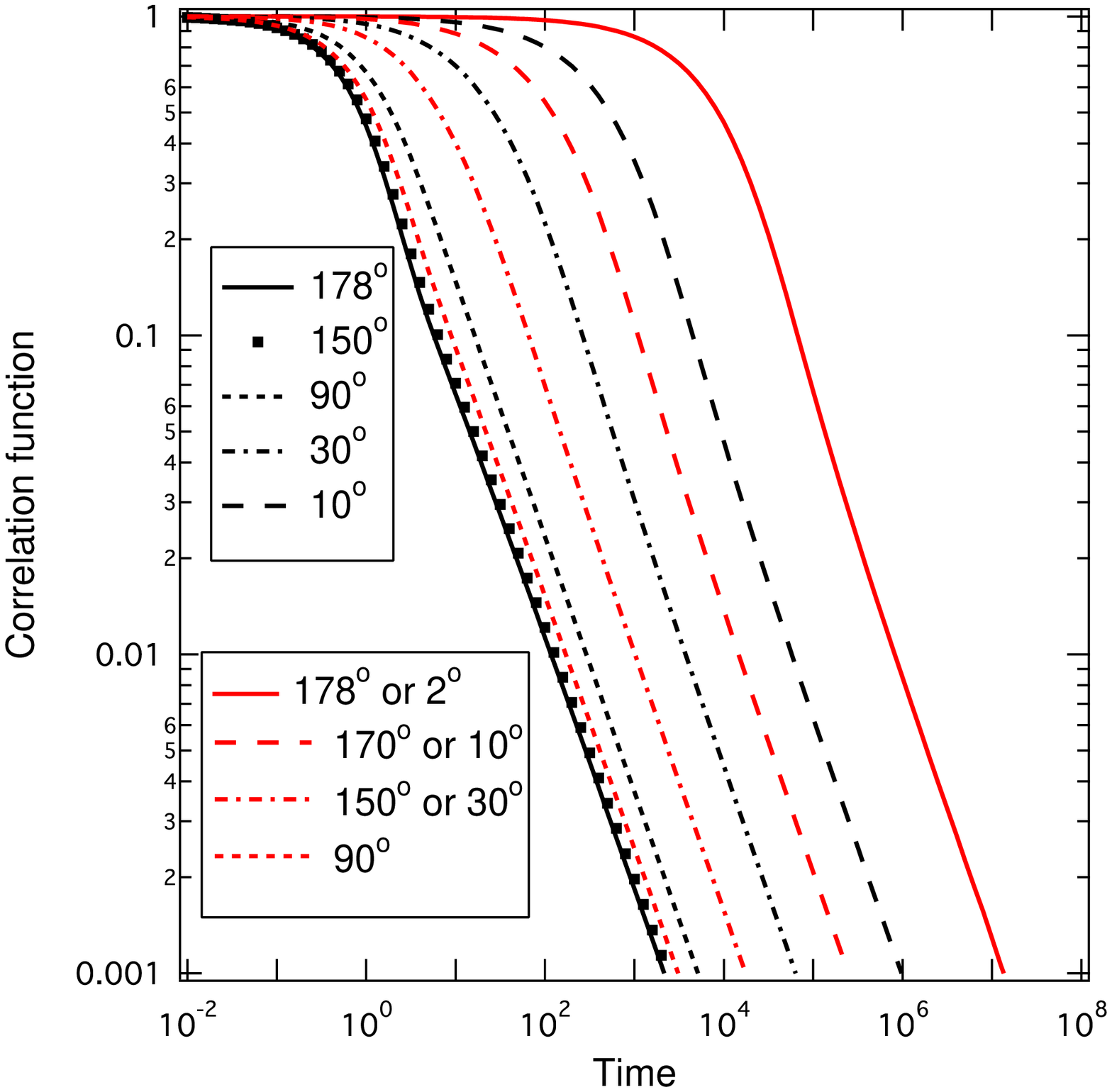}
\caption{}\label{fig:Ivanov4} 
\end{figure}

\newpage
\begin{flushright}
Fig. 6, K. Seki,  B. Bagchi, and M. Tachiya
\end{flushright}
\vspace{5cm}
\mbox{ }\\
\begin{figure}[h]
\includegraphics[width=12cm]{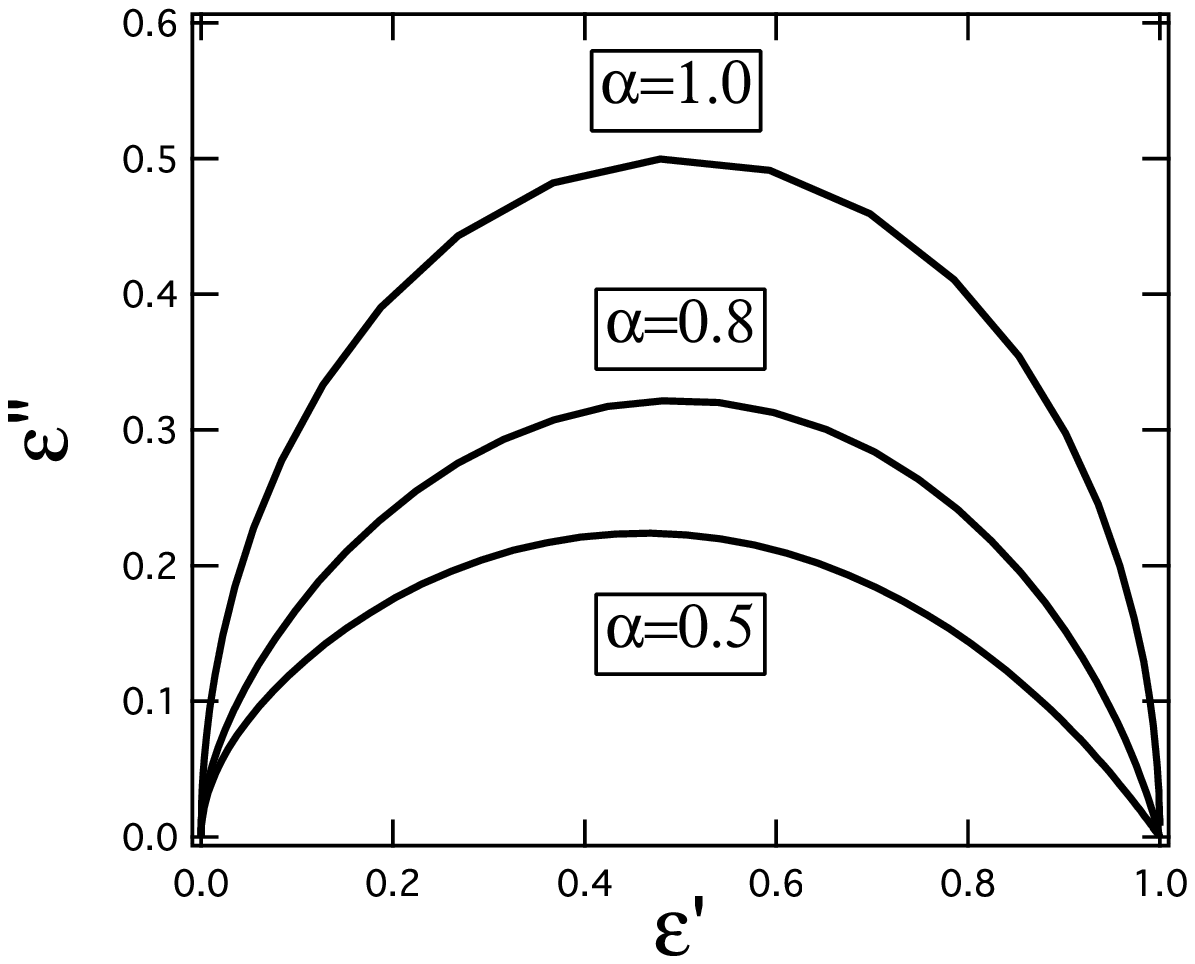}
\caption{}\label{fig:Ivanov5} 
\end{figure}

\newpage
\begin{flushright}
Fig. 7, K. Seki,  B. Bagchi, and M. Tachiya
\end{flushright}
\vspace{5cm}
\mbox{ }\\
\begin{figure}[h]
\includegraphics[width=12cm]{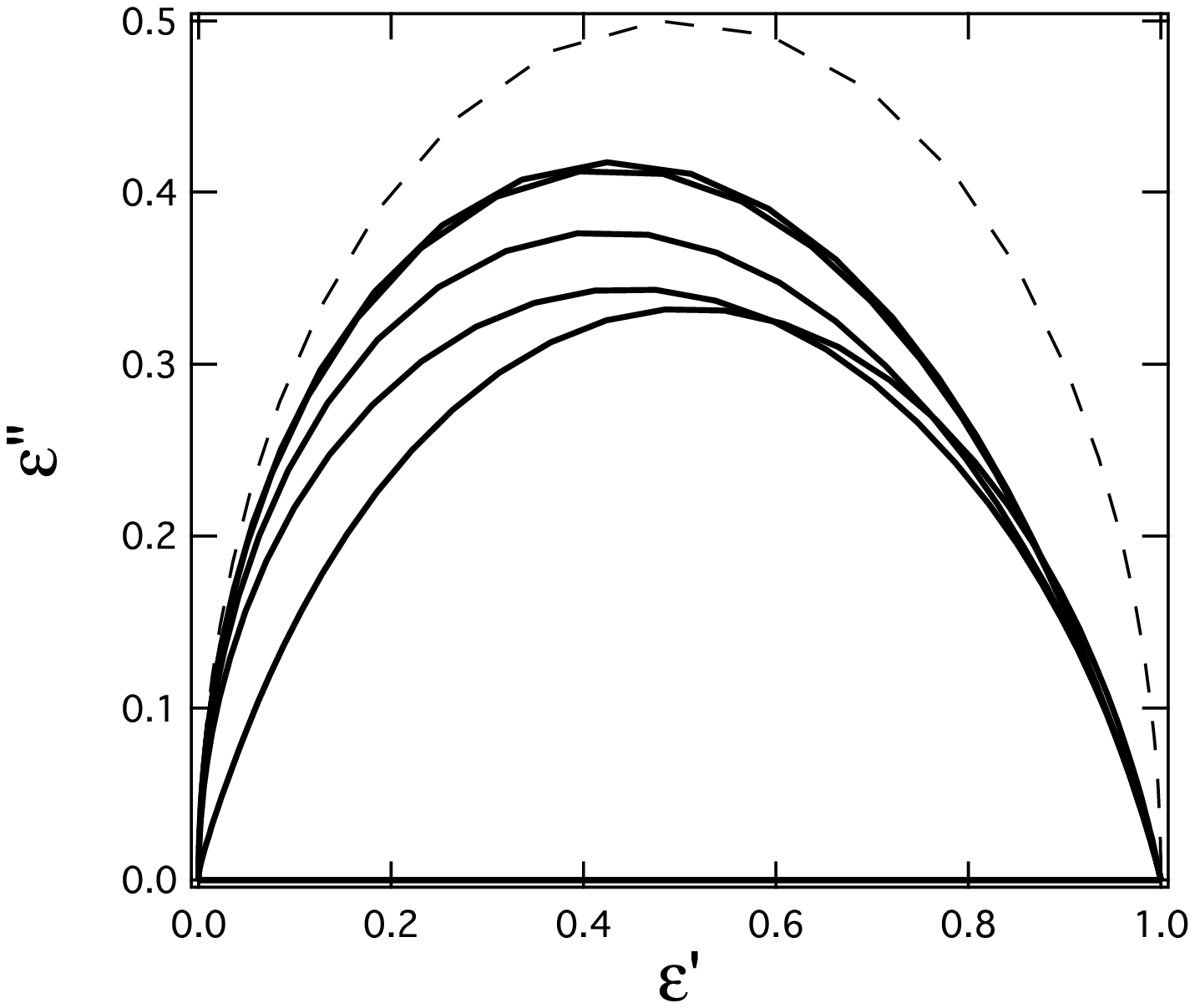}
\caption{}\label{fig:Ivanov6} 
\end{figure}

\newpage
\begin{flushright}
Fig. 8, K. Seki,  B. Bagchi, and M. Tachiya
\end{flushright}
\vspace{5cm}
\mbox{ }\\
\begin{figure}[h]
\includegraphics[width=12cm]{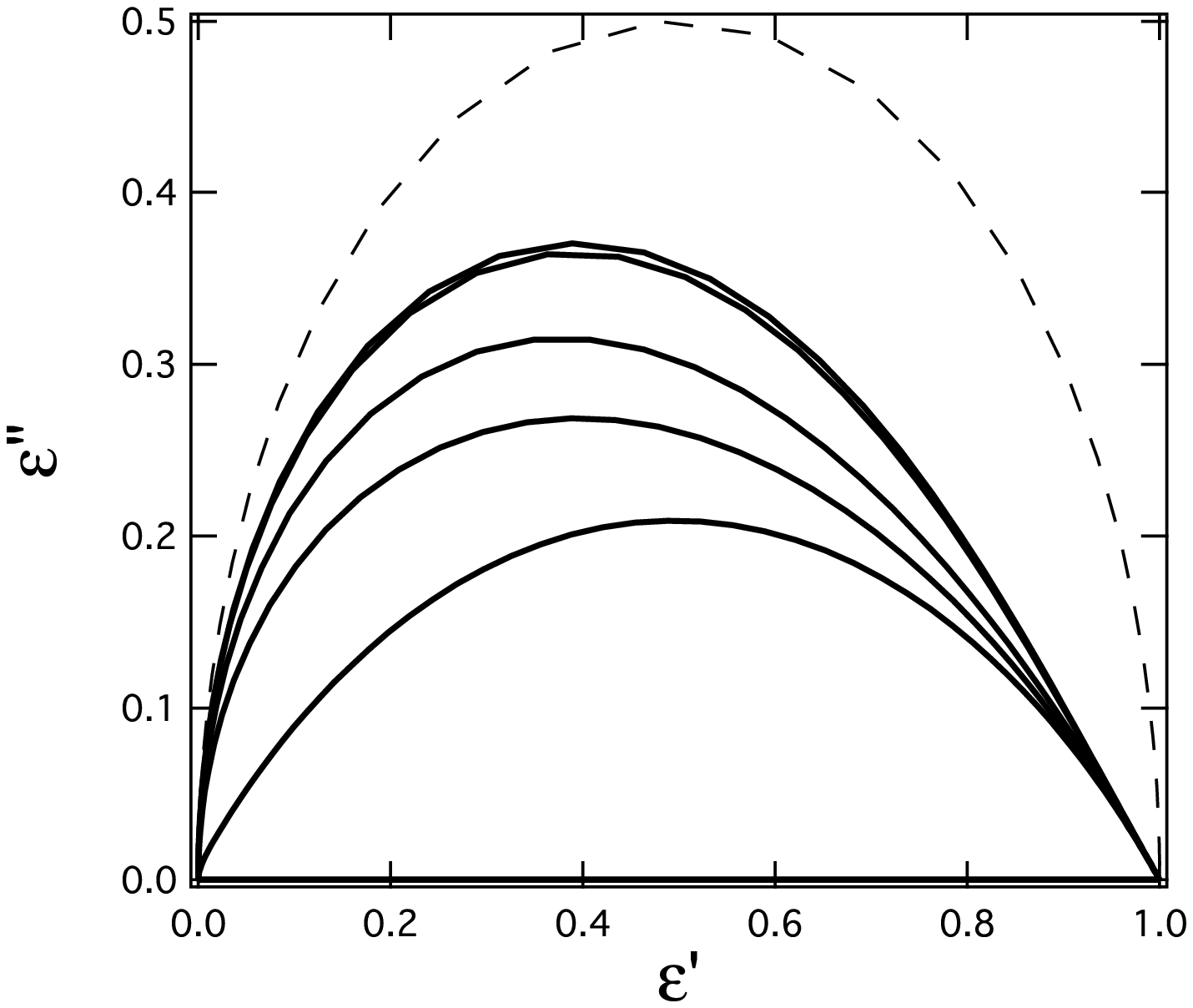}
\caption{}\label{fig:Ivanov7}
\end{figure}

\newpage
\begin{flushright}
Fig. 9, K. Seki,  B. Bagchi, and M. Tachiya
\end{flushright}
\vspace{5cm}
\mbox{ }\\
\begin{figure}[h]
\includegraphics[width=12cm]{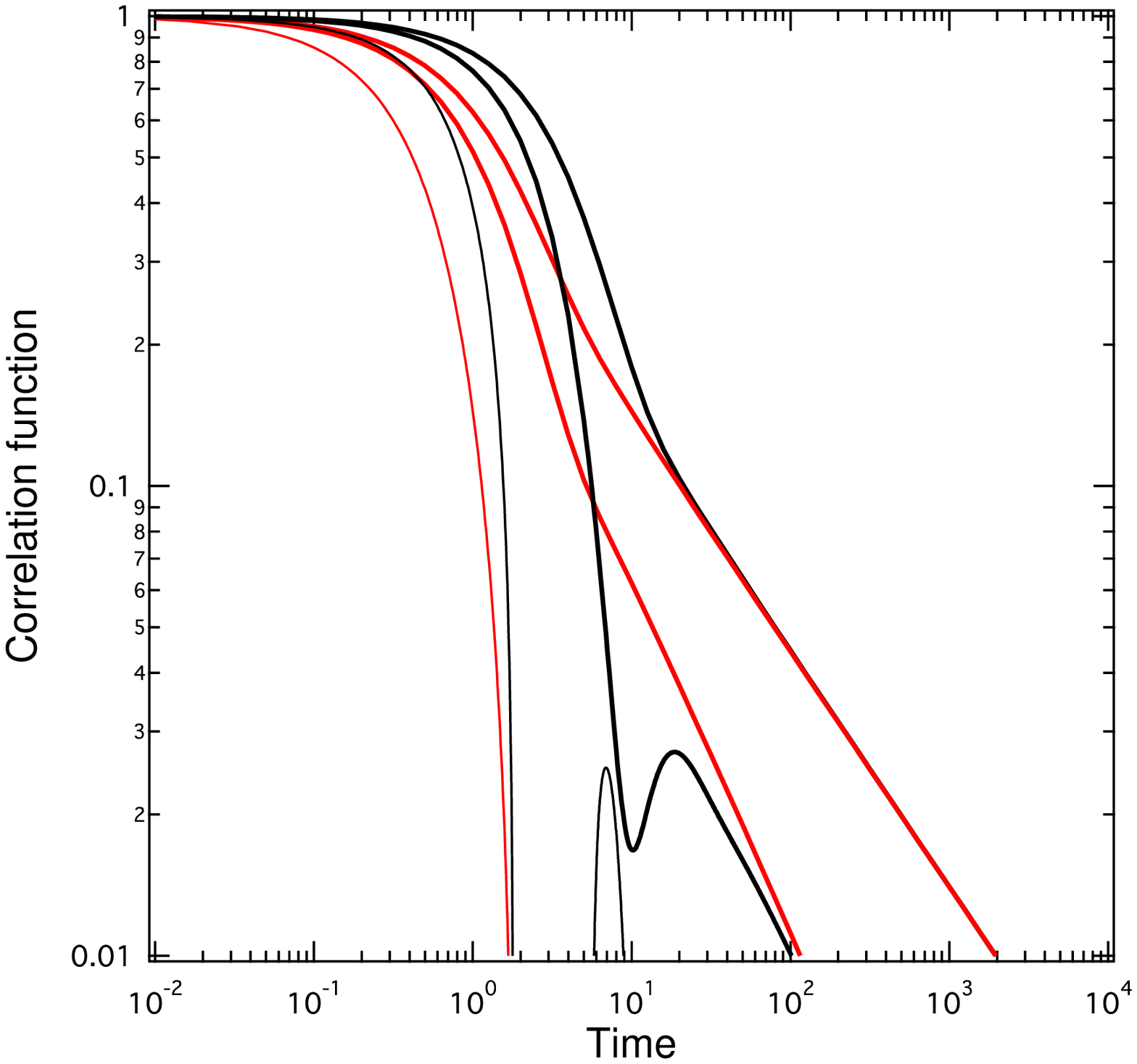}
\caption{}\label{fig:Kubo1} 
\end{figure}

\newpage
\begin{flushright}
Fig. 10, K. Seki,  B. Bagchi, and M. Tachiya
\end{flushright}
\vspace{5cm}
\mbox{ }\\
\begin{figure}[h]
\includegraphics[width=12cm]{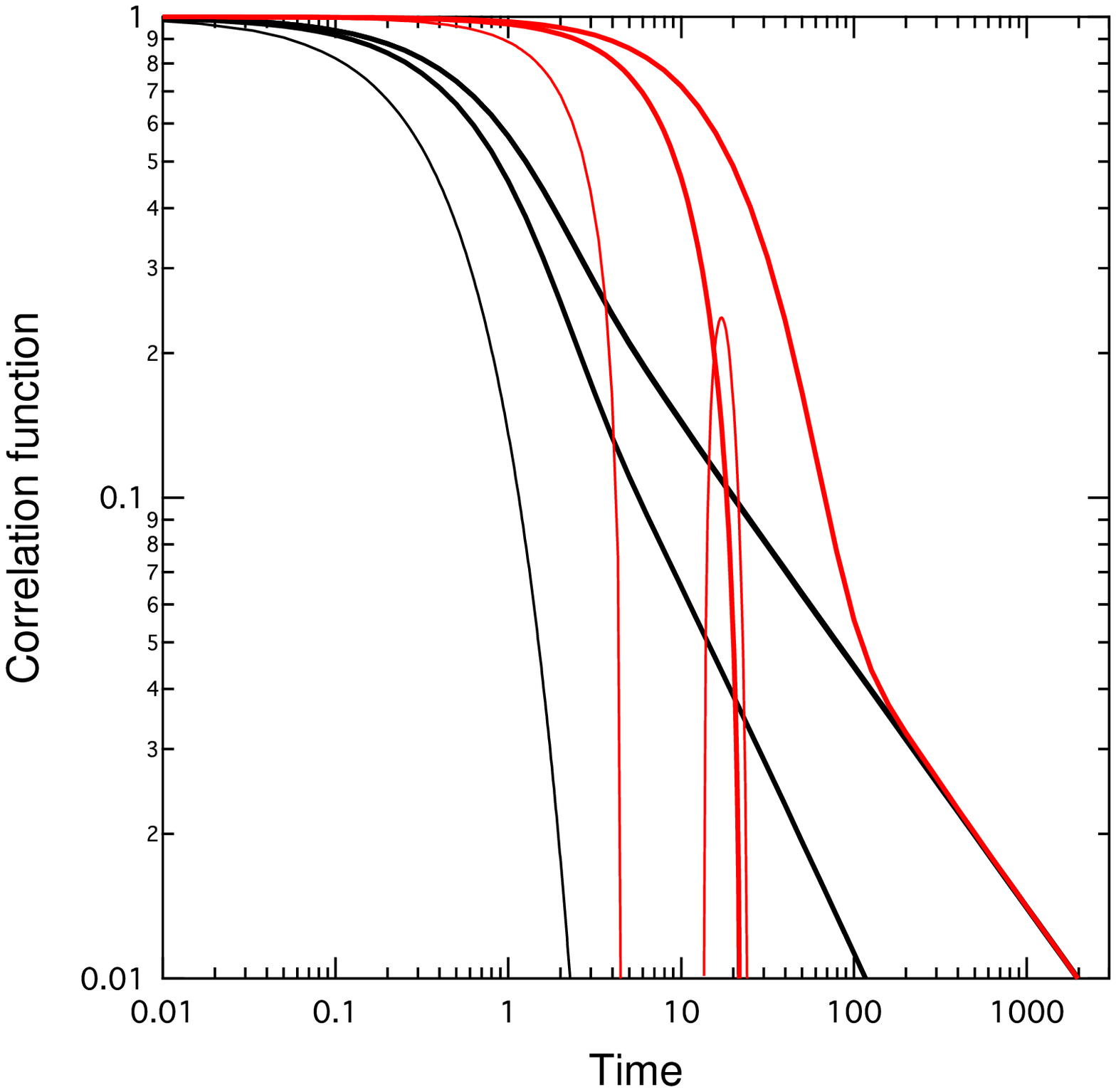}
\caption{}\label{fig:Kubo2} 
\end{figure}

\newpage
\begin{flushright}
Fig. 11, K. Seki,  B. Bagchi, and M. Tachiya
\end{flushright}
\vspace{5cm}
\mbox{ }\\
\begin{figure}[h]
\includegraphics[width=12cm]{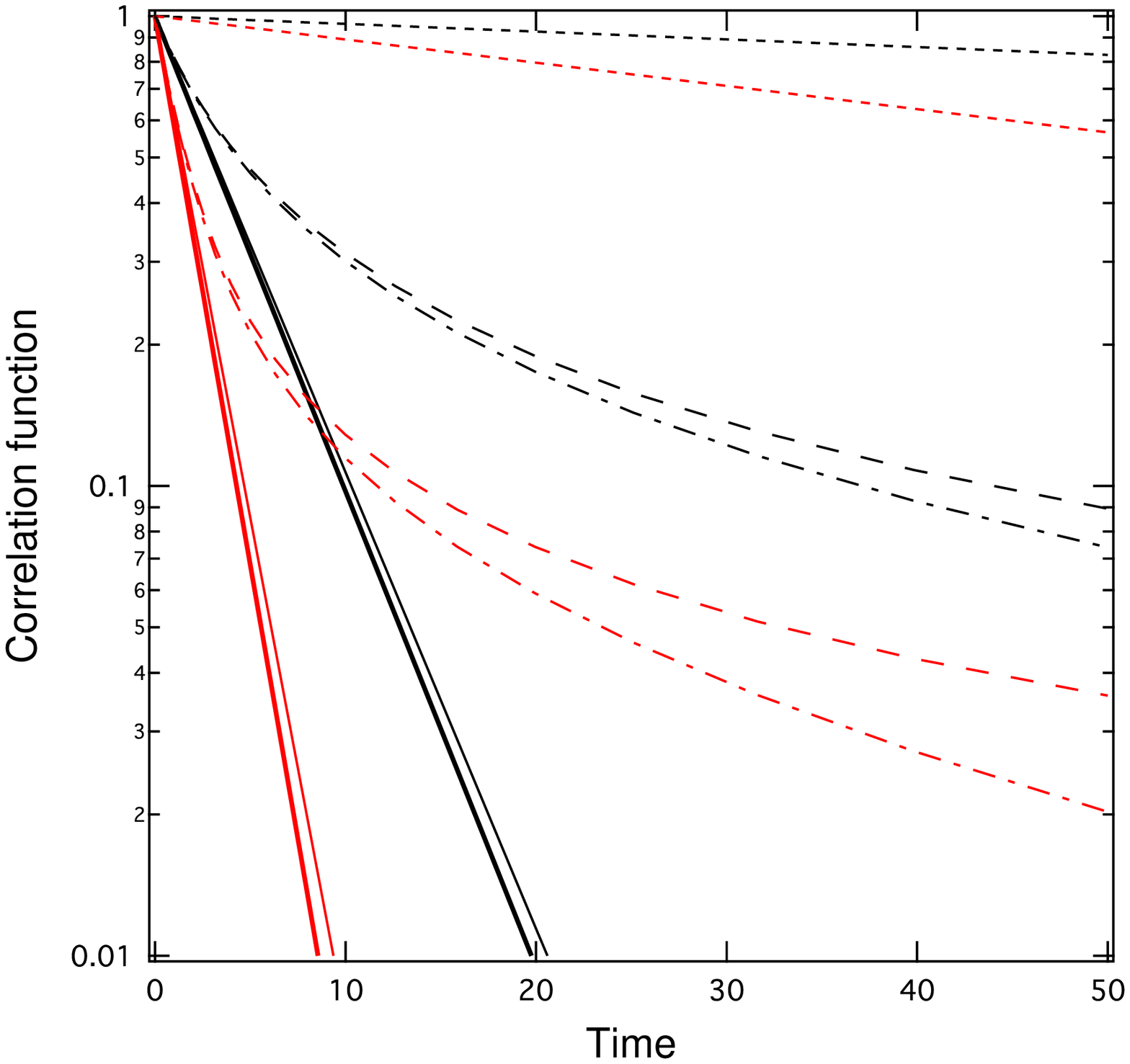}
\caption{}\label{fig:mixed1} 
\end{figure}

\newpage
\begin{flushright}
Fig. 12, K. Seki,  B. Bagchi, and M. Tachiya
\end{flushright}
\vspace{5cm}
\mbox{ }\\
\begin{figure}[h]
\includegraphics[width=12cm]{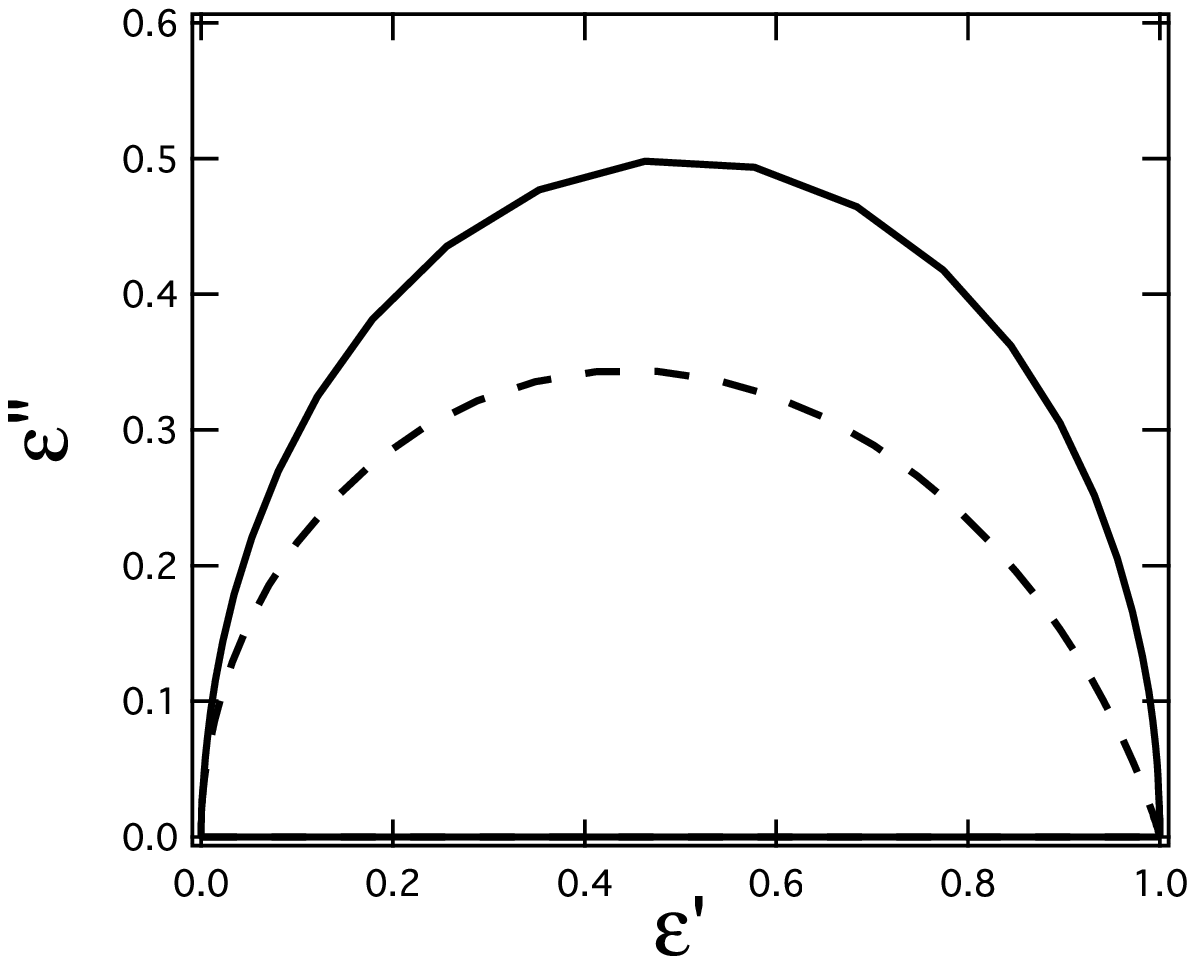}
\caption{}\label{fig:mixed2} 
\end{figure}

\newpage
\begin{flushright}
Fig. 13, K. Seki,  B. Bagchi, and M. Tachiya
\end{flushright}
\vspace{5cm}
\mbox{ }\\
\begin{figure}[h]
\includegraphics[width=12cm]{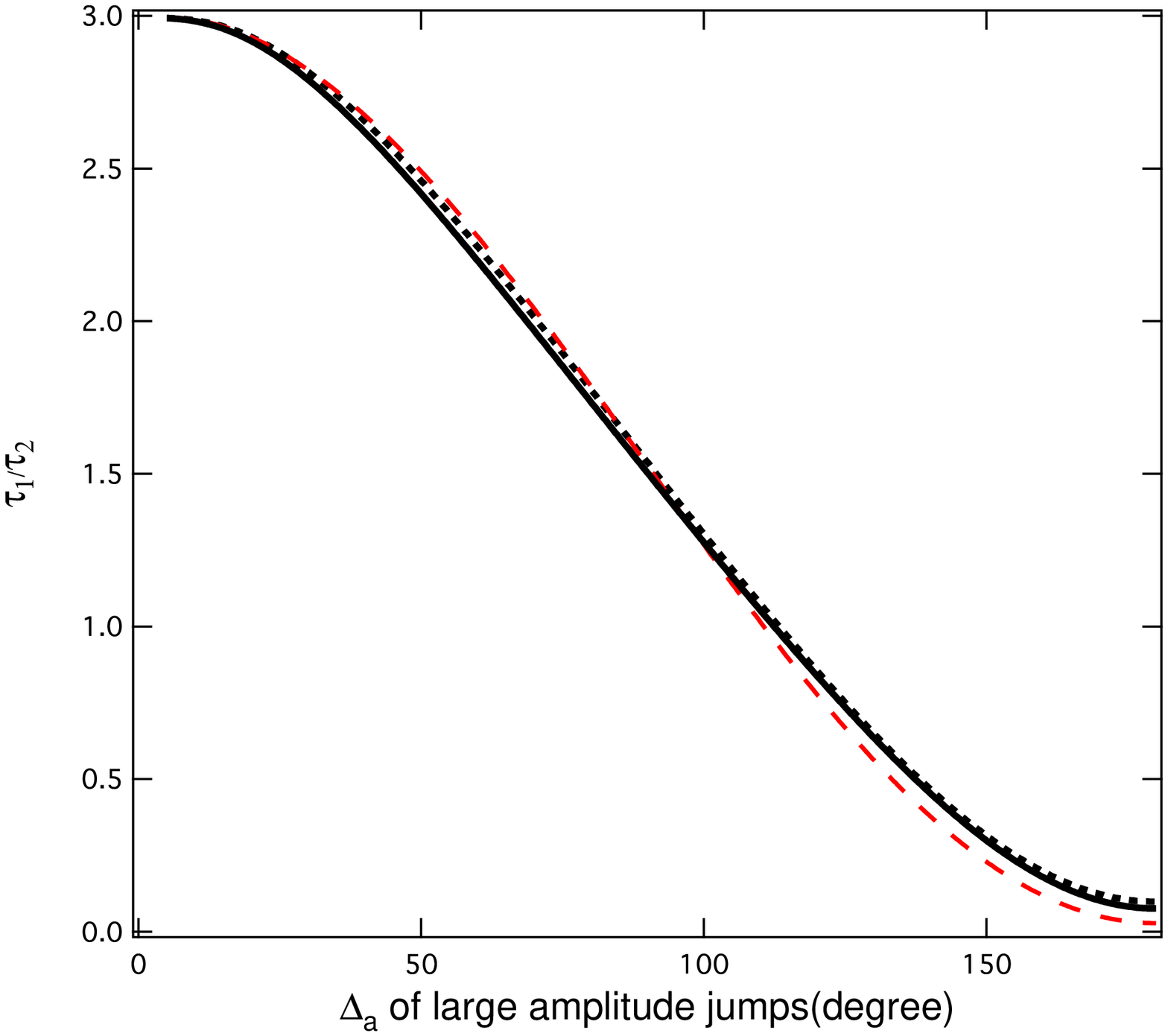}
\caption{}\label{fig:mixed3} 
\end{figure}

\newpage
\begin{flushright}
Fig. 14, K. Seki,  B. Bagchi, and M. Tachiya
\end{flushright}
\vspace{5cm}
\mbox{ }\\
\begin{figure}[h]
\includegraphics[width=12cm]{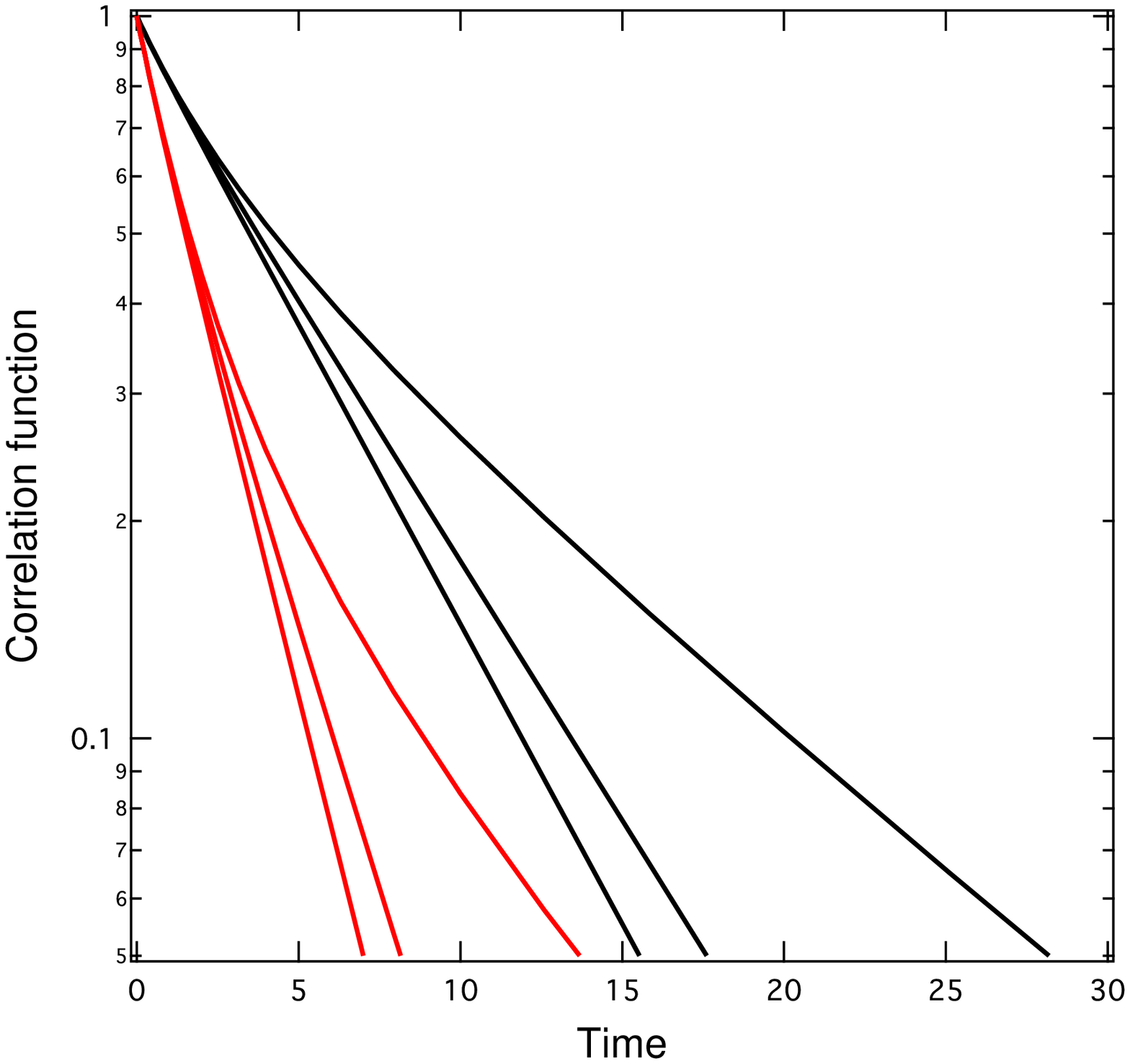}
\caption{}\label{fig:mixed4} 
\end{figure}

\end{document}